\documentclass[pra,onecolumn,superscriptaddress,floatfix]{revtex4-1}
\usepackage{amsmath,amsfonts,amssymb}
\usepackage{graphicx}
\usepackage{epstopdf}
\usepackage[english]{babel}
\usepackage{natbib}
\usepackage[utf8]{inputenc}

\usepackage{letltxmacro}

\LetLtxMacro{\ORIGselectlanguage}{\selectlanguage}
\makeatletter
\DeclareRobustCommand{\selectlanguage}[1]{%
  \@ifundefined{alias@\string#1}
    {\ORIGselectlanguage{#1}}
    {\begingroup\edef\x{\endgroup
       \noexpand\ORIGselectlanguage{\@nameuse{alias@#1}}}\x}%
}
\newcommand{\definelanguagealias}[2]{%
  \@namedef{alias@#1}{#2}%
}
\makeatother

\definelanguagealias{en}{english}
\definelanguagealias{eng}{english}
\definelanguagealias{English}{english}

\usepackage{amsmath,bbold,bbm}
\usepackage{graphicx}
\usepackage{mathtools}
\usepackage[normalem]{ulem}
\usepackage[T1]{fontenc}
\usepackage[utf8]{inputenc}

\let\originalleft\left
\let\originalright\right
\renewcommand{\left}{\mathopen{}\mathclose\bgroup\originalleft}
\renewcommand{\right}{\aftergroup\egroup\originalright}
\renewcommand{\(}{\left(}
\renewcommand{\)}{\right)}
\newcommand{\bra}[1]{\ensuremath{\left\langle #1\right|}}
\newcommand{\ket}[1]{\ensuremath{\left|#1\right\rangle}}

\newcommand{\ie}{\emph{i.e.,\ }}
\newcommand{\eg}{\emph{e.g.,\ }}

\newcommand{\braket}[2]{\left\langle #1,#2\right\rangle}
\newcommand{\<}{\langle}
\renewcommand{\>}{\rangle}
\providecommand{\abs}[1]{\left\lvert#1\right\rvert}
\providecommand{\norm}[1]{\lVert#1\rVert}

\newcommand{\A}{\mathcal{A}}
\newcommand{\h}{\mathfrak{h}}
\newcommand{\x}{\mathfrak{x}}

\renewcommand{\H}{\mathcal{H}}
\newcommand{\1}{\mathbb{1}}

\providecommand{\tr}{{\rm Tr}}
\renewcommand{\phi}{\varphi}
\renewcommand{\equiv}{\coloneqq}
\newcommand\underrel[2]{\mathrel{\mathop{#2}\limits_{#1}}}

\begin{document}
\title{On optimal currents of indistinguishable particles}

\author{Mattia Walschaers} 
\email{mattia.walschaers@lkb.upmc.fr}
\affiliation{Instituut voor Theoretische Fysica, KU Leuven, Celestijnenlaan 200D, B-3001 Heverlee, Belgium}
\affiliation{Physikalisches Institut, Albert-Ludwigs-Universit\"at Freiburg, Hermann-Herder-Str. 3, D-79104 Freiburg, Germany}
\affiliation{Laboratoire Kastler Brossel, UPMC-Sorbonne Universit\'es, CNRS, ENS-PSL Research University, Coll\`ege de France,  4 place Jussieu, F-75252 Paris, France}
\author{Andreas Buchleitner}
\affiliation{Physikalisches Institut, Albert-Ludwigs-Universit\"at Freiburg, Hermann-Herder-Str. 3, D-79104 Freiburg, Germany}
\author{Mark Fannes}
\affiliation{Instituut voor Theoretische Fysica, KU Leuven, Celestijnenlaan 200D, B-3001 Heverlee, Belgium}

\date{\today}

\begin{abstract}
We establish a mathematically rigorous, general and quantitative framework to describe currents of non- (or weakly) interacting, indistinguishable particles driven far from equilibrium. 
We derive tight upper and lower bounds for the achievable fermionic and bosonic steady state current, respectively, which can serve as benchmarks for special cases of interacting many-particle dynamics. For fermionic currents, we identify a symmetry-induced enhancement mechanism in parameter regimes where the coupling between system and reservoirs is weak. This mechanism is broadly applicable provided the inter-particle interaction strength is small as compared to typical exchange interactions.
\end{abstract}

\maketitle

\section{Introduction}

Currents---the specific physical feature of non-equilibrium steady states of open systems subject to a potential gradient established by reservoirs---are a prominent topic of various branches of condensed matter physics. As the sizes of technological devices driven by currents reach mesoscopic scales, non-trivial quantum effects must be taken into account~\cite{steinlesberger_electrical_2002,imry_introduction_2009,moors_modeling_2015,moors_resistivity_2016}. Yet, our theoretical understanding of currents in quantum systems is far from complete, \eg many results are available for perfect lattices~\cite{ashcroft_solid_1976}, but more realistic set-ups, with disorder and decoherence, still pose a panoply of open questions~\cite{derezinski_fluctuations_2008, asadian_heat_2013,manzano_quantum_2012,mendoza-arenas_dephasing_2013, kordas_non-equilibrium_2015, Ref11,manzano_quantum_2016}.

The past decade has seen a vivid debate on the relevance of quantum mechanics in biological systems and most notably in photosynthesis~\cite{engel_evidence_2007,cheng_dynamics_2009,scholes_lessons_2011,scholak_efficient_2011,alicki_resonance_2012,walschaers_optimally_2013,levi_quantum_2015}. Since photosynthetic organisms are immersed in an environment of thermal photons, one may describe the situation via a constant influx of photons triggering an outflow of electrons~\cite{blankenship_molecular_2002}. The system, a large collection of intricately coupled chlorophyll molecules, is therefore constantly experiencing a flow of excitons which may be interpreted as a current. At present, the debate~\cite{jesenko_optimal_2012,jesenko_excitation_2013,manzano_quantum_2013,witt_stationary_2013,shatokhin_coherence_2016} on how such flow in the stationary state can be affected by quantum coherence on transient time scales remains widely open.

Quantum effects do not only emerge naturally in many cases, they can also be engineered. In quantum dots~\cite{reed_observation_1988,beenakker_theory_1991,jalabert_universal_1994,contreras-pulido_time_2012} and molecular junctions~\cite{fulton_observation_1987,nitzan_electron_2003,segal_heating_2002,segal_thermal_2003,velizhanin_heat_2008}, currents have been studied for decades. In addition, cold atom~\cite{chu_cold_2002,kohl_fermionic_2005,ponomarev_atomic_2006,bloch_many-body_2008} and trapped ion~\cite{schneider_experimental_2012} set-ups provide clean testing grounds to study currents in a manifestly quantum mechanical setting, including quantum many-particle and statistical effects.

The aim of this contribution is to provide a rigorous mathematical treatment of currents in non-equilibrium quantum systems. To achieve this goal, we need a model which is analytically controllable. Therefore, we treat the coupling between the system and the particle reservoirs in a Markovian way, \textit{i.e.}\ we ignore memory effects in the dynamics and use therefore a dynamical semi-group. Moreover, we focus on systems in which inter-particle interactions are sufficiently weak, such that the system can be described by effective model of free, \ie~non-interacting particles. For fermions, this implies that the shifts in energy levels associated to inter-particle interactions must be small compare to the energy-level spacings associated with the exchange interaction induced by the exclusion principle. In this scenario, we can derive bounds on the current, which are sufficiently tight to be saturated by properly designed systems. Our results thus also serve as a benchmark for studies of quantum transport in systems where interactions (or other non-linear effects) cannot be ignored. A violation of our analytically derived bounds is an unambiguous indicator of non-trivial interaction-induced effects, beyond mere many-particle interferences between indistinguishable particles.

To establish such a versatile theoretical approach which can handle the above diverse scenarios, and, in particular, also accounts for potential quantum statistical effects on transport, Section \ref{sec:intro} of our present contribution provides a self-contained introduction to the mathematically rigorous framework of many-particle quantum currents. We will herein strongly rely on algebraic quantum statistical mechanics, a formalism which stems from mathematical physics. This algebraic approach to quantum mechanics of many-particle systems is indispensable to study infinitely large systems (as we do in Section \ref{sec:Ribbon}). Within this framework, we introduce a quantum version of the continuity equation, applicable to open system dynamics of the semi-group type~\cite{alicki_detailed_1976,alicki_quantum_1979}. We consider three contributions to the dynamics: a Hamiltonian part for the reversible particle dynamics, and two non-Hamiltonian parts which describe particle injection and extraction, respectively. For such non-equilibrium many-particle systems we derive several fundamental properties: In Sections~\ref{sec:Currents} and~\ref{sec:Bosons} we derive an upper bound for the particle current in the fermionic setting, and a lower bound for bosonic systems, respectively. In Section~\ref{sec:Symmetry}, we show that the fermionic upper bound can be saturated by appropriate design of the Hamiltonian part of the dynamics. The algebraic framework allows us to go beyond the standard Fock space formalism, which we illustrate in Section~\ref{sec:Ribbon}, where we derive an upper bound for the current density in a ribbon, \ie a  2D lattice system with shift-invariance in one direction, and a finite width in the other.

The strength of our contribution is that it makes no assumptions on the underlying single-particle Hamiltonian, and that it is applicable whenever the interaction between particles can be ignored to a good approximation. Hence, our approach does not only provide fundamental insight on the achievable currents in non-equilibrium quantum systems, but also opens novel perspectives for research in the fields mentioned above, where one may exploit the here identified design principle in a specific context.

\section{Many-fermion systems}\label{sec:intro}

We first provide an introduction to the algebraic formalism which describes many-fermion systems. The results presented in this section are well-known in the mathematical physics literature on quantum statistical mechanics~\cite{alicki_field-theoretical_2010,alicki_quantum_2001,benatti_quantum_2010,bratteli_operator_1997}. In Sections \ref{sec:Currents} to \ref{sec:Bosons}, we apply this formalism to investigate the physics of currents in open quantum systems.
 
\subsection{Fock space}\label{sec:FockSpace}
It is common practice to describe many-fermion systems in terms of Fock space. This space is formally constructed using a single-particle Hilbert space $\H$, also referred to as the mode space, as basic building block which provides all degrees of freedom of a single particle. As postulated by Pauli, identical particles are independent of labelling, a constraint which either leads to bosons or fermions. The wave functions of the latter species change sign under odd permutations of particles which is reflected in the fermionic $n$-particle Hilbert space 
\begin{equation}\label{eq:FockFerm1}
\H^{(n)} = \H \otimes \H \otimes \cdots \otimes \H \Big\vert_{\rm asym} \, .
\end{equation}
The anti-symmetrisation implies that the space $\H^{(n)}$ is linearly generated by functions of the form
\begin{equation}
\label{sla}
\psi_1 \wedge \psi_2 \wedge \cdots \wedge \psi_n \equiv \frac{1}{\sqrt{n!}} \sum_{\pi \in S_n} {\rm sign}(\pi)\, \psi_{\pi(1)} \otimes \psi_{\pi(2)} \otimes \cdots \otimes \psi_{\pi(n)} \, .
\end{equation}
Here $S_n$ denotes the permutation group of $n$ objects, $\pi$ a permutation, and $\mathrm{sign}(\pi)$ the signature of $\pi$. Note that these functions are generally not normalised and that they vanish whenever the single-particle wave functions are linearly dependent, as expected from fermions. Functions of the type~(\ref{sla}) are often called Slater determinants. 

The fermionic Fock space $\Gamma(\H)$ constructed on $\H$ is built to accommodate any number of particles and therefore glues together all $n$-particle spaces:
\begin{equation} \label{eq:FockFerm3}
\Gamma(\H) = \mathbb{C} \oplus \H \oplus \H^{(2)} \oplus \H^{(3)} \oplus \cdots,
\end{equation}
where $\mathbb{C}$ describes the vacuum component where no particles are present in the system.
In the fermionic case and for a finite dimensional mode space $\H$ the direct sum breaks off: Fermionic Fock space can never harbour more particles than $\dim(\H)$. Often it is assumed that Fock space is sufficient to describe general many-particle systems which is slightly inaccurate. Fock space can only accommodate finite numbers of particles. Both in the case where $\H$ is not finite dimensional, \eg for infinitely extended systems (see Section \ref{sec:Ribbon}),  or for a bosonic system (see Section \ref{sec:Bosons}), physics is much richer than Fock space. To study this larger realm of many-particle quantum physics, we must switch to a description in terms of observables and select a Hilbert space representation that matches the given physical situation.

\subsection{Algebra of observables}

The main tools at hand in Fock space are the creation and annihilation operators: $a^{\dag}(\phi)$ and $a(\phi)$ respectively, where $\phi \in \H$. We work in a formalism of non-local creation and annihilation operators which have a straightforward interpretation: they create and annihilate a single-particle state $\phi$. Their action is easily given on Slater determinants: 
\begin{equation}\label{cre:CAR}
a^{\dag}(\phi)\, \psi_1 \wedge \psi_2 \wedge \cdots \wedge \psi_n = \phi \wedge \psi_1 \wedge \psi_2 \wedge \cdots \wedge \psi_n \, ,
\end{equation}
where we have identified the $n$-particle vector $\psi^{(n)} =  \psi_1 \wedge \psi_2 \wedge \cdots \wedge \psi_n \in \H^{(n)}$ with $0 \oplus 0 \oplus \cdots \oplus \psi^{(n)} \oplus \cdots$ in $\Gamma(\H)$. The annihilation operator is the adjoint of the creation operator, its action on a Slater determinant is
\begin{equation}\label{anh:CAR}
a(\phi)\, \psi_1 \wedge \psi_2 \wedge \cdots \wedge \psi_n = \sum_{j=1}^n (-1)^{j+1}\, \braket{\phi}{\psi_j}\, \psi_1 \wedge \dots \wedge \psi_{j-1} \wedge \psi_{j+1} \wedge \cdots \wedge \psi_n \, .
\end{equation}
Indeed, as one may expect for fermions, sign bookkeeping is required. 

Fermionic creation and annihilation operators obey the {\em canonical anti-commutation relations} (CAR)
\begin{equation}\label{eq:CAR1}
\{a(\phi), a^{\dag}(\psi)\} = \braket{\phi}{\psi}\, \1
\quad{\rm and}\quad
\{a(\phi), a(\psi)\} = 0\qquad \forall \, \psi, \phi \in \H \, .
\end{equation}
These operators generate an algebra that forms the basic mathematical framework for the description of many-fermion systems with a given mode space $\H$. The key idea of algebraic quantum physics is that the {\em algebra of observables}, rather than a Hilbert space, is the central mathematical object to describe large quantum systems.

As a general algebraic framework and to contrast it with the Fock space representation above, we introduce {\em abstract creation and annihilation operators} $c^*(\psi)$ and $c(\phi)$ respectively, $\phi, \psi \in \H$. It must be emphasised that these objects are no longer linear operators on the Fock space, but merely generate a formal algebra determined by the basic relations
\begin{align}
&\psi \in \H \mapsto c^*(\psi) \quad \text{is $\mathbb{C}$-linear} \\
&\{c(\phi), c^*(\psi)\} = \braket{\phi}{\psi}\, \1
\quad\text{and}\quad
\{c(\phi),c(\psi)\} = 0 \, \label{eq:CAR}.
\end{align}
The $\ast$ is a formal operation which is the abstract version of the Hilbert space adjoint $\dagger$. One then completes the algebra with respect to the unique $C^*$-norm to obtain the $C^*$-algebra $\A^{\text{CAR}}$ of the CAR on $\H$.\footnote{A $C^*$-algebra $\mathcal{A}$ is by definition equipped with a norm which fulfils the properties $\norm{ x^* } = \norm{ x }$ and $\norm{x^* x} = \norm{ x^*} \norm{ x}$ for all $x \in \mathcal{A}$. Here, this demand is strong enough to fix the norm in a unique way which is why it is referred to as the $C^*$-norm. For a much more complete and formal introduction to $C^*$-algebras, see for example \cite{bratteli_operator_1987}.} The completion is needed to apply general mathematical results and to describe dynamics in a controlled way. This framework is necessary to describe general many-particle systems with infinite-dimensional single-particle spaces; in these systems, we cannot describe all possible physics for all possible states (see Section \ref{sec:states}) on the level of Fock spaces. In our present contribution, we strictly require this framework for the study of the quantum ribbon in Section \ref{sec:Ribbon}. 

In this formalism, observables are those objects $O \in \A^{\rm CAR}$ which are constructed using $\1$, $c^*(\psi)$ and $c(\phi)$ and which have the additional property that $O = O^*$. In the context of many-particle systems, it is often useful to focus on polynomials of $c^*$ and $c$, in which contributions with specific particle numbers are related to definite orders. In this work, we focus solely on the simple class of single-particle observables corresponding to polynomials of order two.

A single-particle observable is essentially an embedding of an observable on the one-particle space into the many-particle framework. In Fock space, one assigns a copy of the observable to each different particle in an additive way, \eg the total energy of a system described by a single-particle Hamiltonian is the sum of the single-particle operators for each separate particle. The formal algebraic way to express this {\em second quantisation} is via the mapping $\Gamma: \mathcal{B}(\H) \rightarrow \A^{\rm CAR}$ from the space of bounded operators on $\H$ to the algebra of observables, which acts as
\begin{equation}\label{quant:sec}
\Gamma(O) \equiv \sum_{i,j} \braket{\eta_i}{O\, \eta_j} c^*(\eta_i)\, c(\eta_j)
\end{equation} 
where we may select any orthonormal basis $\{\eta_j\}$ of $\H$. In order to ensure that $\Gamma(O)$ belongs to the algebra one has to impose the rather restrictive condition that $O$ is a trace-class operator. It is not hard to check that different bases $\{\eta_j\}$ yield a same second quantised observable.

A specific example of interest in the discussion of particle currents for finite dimensional one-particle spaces $\H$ is the {\em number operator} $N$ which literally counts the number of particles in the system. This operator is in essence of single-particle type, as it is given by
\begin{equation}
N \equiv \Gamma(\1) = \sum_j c^*(\eta_j)\, c(\eta_j)\, .
\end{equation}
Indeed, particle currents describe the in- and outflow of particles and therefore the behaviour of the observable $N$ goes hand in hand with the behaviour of such currents.

Not only does the algebraic formalism require a more abstract description of the observables in our theory, it also implies a more general structure for the quantum states which determine the statistics of measurement outcomes for these observables.

\subsection{States}\label{sec:states}

A quantum state is commonly associated either with a state vector $\psi$ or with a density matrix $\rho$. Expected values of observables $O$ are given by $\<O\> = \<\psi, O\,\psi\>$ or $\<O\> = \tr \rho\, O$. This presupposes a specific Hilbert space representation of the physical system. The more general algebraic formalism starts with expectation functionals that allow for a probabilistic interpretation~\cite{davies_operational_1970, holevo_statistical_2001,maassen_quantum_2010}. Thus a state is a linear functional $\omega: \A^{\rm CAR} \to \mathbb{C}$ on the algebra of observables fulfilling the requirements
\begin{equation}
\omega(\1) = 1
\quad\text{and}\quad
\omega(x^*x) \geqslant 0 \qquad \forall x \in \A^{\rm CAR}\, .
\end{equation}
These properties are respectively known as the {\em normalisation} and {\em positivity} conditions. 

A useful tool to describe states, and their perturbations, on a C*-algebra is the Gelfand-Naimark-Segal (GNS) construction \cite{alicki_quantum_2001,bratteli_operator_1997,bratteli_operator_1987,verbeure_many-body_2011}. This procedure associates a unique, but state-dependent, Hilbert space representation of the algebra to state $\omega$. This representation returns the state as an expectation with respect to a state vector. Different states may, however, lead to inequivalent representations, which typically happens in the thermodynamic limit of many particle-systems. As an example one may consider Bardeen-Cooper-Schrieffer theory \cite{PhysRev.108.1175,Bogolubov1960S1,Haag1962,doi:10.1063/1.1931227,Thirring1967,Thirring1968,Balslev1968,Goderis1991}, where states with a finite particle density in the thermodynamic limit must be represented in a different Hilbert space than the Fock space which is constructed by exciting the physical vacuum (see Section \ref{sec:FockSpace}). The GNS construction is a key result in algebraic quantum physics, which stresses that the properties of the system's state are essential prerequisites to study physical models.

States on $\A^{\text{CAR}}$ are usually characterised in terms of {\em correlation functions}, \ie one strives to define all objects of the form 
\begin{equation}\label{eq:corr}
\omega\Big( c^*(\psi_1) \cdots c^*(\psi_m)\, c(\phi_n) \cdots c(\phi_1) \Big)\, .
\end{equation}
In the present context, where only single-particle observables are considered, there are simple ways to describe the relevant expectation values. A notable fact is the existence of a linear operator $Q \in \mathcal{B}(\H)$ for each $\omega$, which serves as a (non-normalised) density matrix and is commonly interpreted as a covariance matrix: 
\begin{equation}\label{eq:Q}
\omega\big(c^*(\psi)\, c(\phi)\big) = \braket{\phi}{Q \, \psi}\, .
\end{equation}
In the class of gauge-invariant quasi-free states, this operator $Q$ suffices to fully determine the state.\footnote{An accessible introduction to fermionic quasi-free states can be found in~\cite{dierckx_fermionic_2008}.} In general, this is far from true and one can just say that $Q$ characterises the single-particle statistics. 

The fact that we are considering states on the CAR-algebra directly implies that $0 \leqslant Q \leqslant \1$. The first inequality is necessary to fulfil positivity of the state, the second is a consequence of the fermionic behaviour and represents Pauli's exclusion principle. It follows~\cite{bratteli_operator_1997}
that for a general single-particle observable $\Gamma(B) \in \A^{\rm CAR}$, with $B$ a trace-class operator on $\H$,
\begin{equation}
\label{eq:expB}
\omega\big(\Gamma(B) \big) = \tr (Q\,B) \, .
\end{equation}
This identity might not seem spectacular but it offers an enormous computational simplification. It is, therefore, one of the key ingredients in all the following sections of the present contribution.

If $\omega$ is a normal\footnote{For an operator algebra of observables which acts on a Hilbert space, a state is said to be {\em normal} if it can be represented by a density operator which is a trace-class operator on the same Hilbert space. However, because we consider abstract C*-algebras, it only makes sense to refer to {\em normal} states in the context of a specific representation. Throughout our contribution we will always refer to {\em normal} states as states which can be represented by a density matrix on the {\em Fock space}, in other words, states which are normal with respect to the Fock representation. Note that in this representation the abstract operators $c$ are represented by the operators $a$ of (\ref{eq:CAR1}).} gauge-invariant quasi-free
state it can be shown that
\begin{equation}
\omega (N) = \tr\, Q\, < \infty\, ,
\end{equation}
hence directly expressing the expected particle number in terms of $Q$ which is now also a trace-class operator. This condition is also sufficient to guarantee normality.

\subsection{Dynamics}
\label{sec:dyn}

In the spirit of the algebraic approach, it makes sense to consider the elements of the algebra $\A^{\rm CAR}$ as the dynamical objects in the theory, whereas the states remain unchanged at all times. This formally implies that we can consider a mapping $\Lambda_{t_1,t_0}: \A^{\rm CAR} \rightarrow \A^{\rm CAR}$ for an evolution from time $t_0$ up to $t_1$. The first obvious requirements for a well-defined dynamics are
\begin{equation}\label{eq:dynMapCond}
\Lambda_{t_1,t_0}(\1) = \1
\quad \text{and}\quad
\Lambda_{t_1,t_0}(x^*x) \geqslant 0 \quad \forall x \in  \A^{\rm CAR}\, .
\end{equation}
These demands must be fulfilled for any choice of $t_1$ and $t_0$. 

A more debatable~\cite{pechukas_reduced_1994,alicki_comment_1995,pechukas_pechukas_1995} assumption on the dynamics is {\em complete positivity} which formally says that the system can be trivially embedded in a larger system without having to fear for loss of positivity. Such embeddings are also important to include internal degrees of freedom in the description. Complete positivity in other words guarantees that effective descriptions of only a subset of relevant degrees of freedom are possible. The formal mathematical phrasing requires an extension of the algebra by any matrix algebra ${\cal M}_N$ to obtain $\A^{\rm CAR}\otimes {\cal M}_N$. We may now trivially extend $\Lambda_{t_1,t_0}$ on $\A^{\rm CAR}$ to $\Lambda_{t_1,t_0} \otimes {\rm id}_N$ on $\A^{\rm CAR}\otimes {\cal M}_N$. When $\Lambda_{t_1,t_0} \otimes {\rm id}_N$ is a positive map for {\em any} $N$, the dynamics is said to be completely positive~\cite{stinespring_positive_1955, kraus_general_1971,lindblad_generators_1976}.

In addition to complete positivity, one may impose another demand which rarely holds exactly for a real physical system but often provides a very good approximation~\cite{davies_quantum_1976, alicki_quantum_1987,breuer_theory_2007}: We impose a one-parameter semi-group structure on our dynamical map. The term ``semi-group'' implies divisibility of the map and hence the existence of a {\em generator}. Moreover, the generator is time-independent and thus the map is only governed by $t = t_1 - t_0$. In other words, we can write the dynamics in terms of $\Lambda_t$, and obtain that
\begin{equation}
\Lambda_{t+s} = \Lambda_t \circ \Lambda_s = \Lambda_s \circ \Lambda_t \quad \forall t,s \geqslant 0\, .
\end{equation}  
In general, we do not assume that the inverse exists, thus withholding the family of maps from being a full-blown group. 

This type of dynamics is particularly useful due to powerful mathematical results. The results by Gorini, Kossakowski, Sudarshan~\cite{gorini_completely_1976} and Lindblad~\cite{lindblad_generators_1976} are well-know, but only hold for algebras of observables which can be described by bounded operators on a Hilbert space. Nevertheless, Lindblad provided a more general recipe for completely positive, one-parameter semi-group dynamics on a C*-algebra $\A$: He showed that any equation of motion of the type 
\begin{equation}\label{eq:general}
\begin{split}
&\frac{\rm d}{{\rm d} t} x =  \Psi(x) +k\,x+x\, k^* \quad  \forall x \in \A\, ,\enskip\text{with} \\
&k \in \A \enskip\text{and}\enskip \Psi: \A \rightarrow \A \, \text{ a completely positive map}\, ,
\end{split}
\end{equation}
leads to a dynamical map with such properties. Hence, we may follow this prescription to engineer a dynamical system with the desired phenomenological properties. In other words, we do not microscopically derive a master equation but rather study one which has the correct phenomenology.

In our present work, we follow and explore a model described by Davies~\cite{davies_irreversible_1977}. From here onward, we assume that $\H$ is finite dimensional which is a considerable technical simplification. In section~\ref{sec:Ribbon}, however, we will deal with translation-invariant systems and discuss how to cope with this more general situation. In particular, (\ref{eq:general}) allows us to write the generator of the dynamical semi-group in a form that nearly resembles the standard Lindblad form \cite{davies_irreversible_1977}: 
\begin{align}\label{eq:Lindblad1}
&\frac{\rm d}{{\rm d} t} x = i[\h,x] + \mathcal{D}^d(x) + \mathcal{D}^a(x),\\
&\mathcal{D}^d(x) := \sum_j \gamma^d_j \( c^*(\delta_j)\theta(x) c(\delta_j) - \frac{1}{2}\, \left\{c^*(\delta_j)c(\delta_j), x \right\}\) \label{eq:Lindblad2}\\
&\mathcal{D}^a(x) := \sum_j \gamma^a_j \( c(\alpha_j)\theta(x) c^*(\alpha_j) - \frac{1}{2}\, \left\{c(\alpha_j)c^*(\alpha_j), x \right\}\), \label{eq:Lindblad3}
\end{align}
where $\alpha_j, \delta_j \in \H$. To make sure that $\Psi$ in (\ref{eq:general}) is a CP-map, one must impose \cite{davies_irreversible_1977} that $\theta$ in~(\ref{eq:Lindblad1}) is the $\ast$-automorphism determined by
\begin{equation}
\theta\big(c^*(\psi)\big) = -c^*(\psi),\quad \psi \in \H\, .
\end{equation}
The Lindblad generators $\mathcal{D}^a$ and $\mathcal{D}^d$ describe the injection and extraction of particles into and from the system, respectively. With respect to the system degree of freedom, these terms mediate absorption and dissipation, thus the superscripts $a$ and $d$. More specifically, fermions described by single-particle state vectors $\{\alpha_j\}$ are injected into the system with positive rates $\{\gamma_j^a\}$, and particles which state vectors $\{\delta_j\}$ are lost from the system with positive rates $\{\gamma_j^d\}.$ Note that also temperature dependences can be accommodated within the positive rates $\{\gamma^{a/d}_j\}.$ We consider systems of non-interacting particles, therefore, in accordance to (\ref{quant:sec}), we must set $\h = \Gamma(H)$ with $H=H^{\dag} \in \mathcal{B}(\H).$

We follow the model of~\cite{davies_irreversible_1977}, and many results of that paper are relevant for the present one. Specifically, we are interested in the dynamics of single-particle observables, given by $x = \Gamma(B)$ with $B \in \mathcal{B}(\H)$. Using (\ref{quant:sec}), we insert $\Gamma(B)$ into (\ref{eq:Lindblad1}) followed by a straightforward computation \cite{alicki_quantum_1987,davies_irreversible_1977, alicki_theory_1978} based on the anti-commutation relations (\ref{eq:CAR}), and we find that the relevant equation of motion for one-particle observables is given by
\begin{equation}\begin{split}\label{eq:dyn1}
&\frac{\rm d}{{\rm d} t} \Gamma(B) = i [\Gamma(H), \Gamma(B)] - \{\Gamma(P), \Gamma(B) \} +2\, \tr \big(AB\big) \1\, ,\text{with} \\
&P = A + D\, ,\enskip A \equiv  \sum_{j=1}^{n_a} \frac{\gamma^a_j}{2} \ket{\alpha_j}\bra{\alpha_j}\, ,\enskip\text{and}\enskip D \equiv  \sum_{k=1}^{n_d} \frac{\gamma^d_k}{2} \ket{\delta_k}\bra{\delta_k}\, .
\end{split}
\end{equation}
That $P,A,D \in \mathcal{B}(\H)$ directly follows from their definitions. Because the semi-group dynamics generated by (\ref{eq:dyn1}) is Markovian, all rates must be positive, which in turn implies that $A\geqslant 0,$ $D \geqslant 0$, and hence $P \geqslant 0$. Moreover, for convenience in Section \ref{sec:ProofBound}, we assume that $A$ and $D$ are strictly positive. For the bound on the current that will be derived later on, we can always consider the general case $A\geqslant 0$ and $D\geqslant 0$ using continuity.   

\subsection{Non-equilibrium steady states}\label{sec:NESS}

Now that the equations of motion are determined, we observe that they can be solved exactly:
\begin{equation}\label{eq:GenSol}
\Lambda_t \big( \Gamma (B) \big) = \Gamma \Big(e^{-(P - iH) t } B e^{- (P+iH) t } \Big) + 2\,\int_0^t {\rm d}s\, \tr\Big(e^{-(P - iH) s } B e^{- (P+iH) s } A \Big)\1\, .
\end{equation}
We notice that, through its dependence on the absorption generator $A$, the second term is specifically related to the population of the system, via the particles that are pumped in. To infer the statistical distribution of measurement results associated with the observable $\Lambda_t(\Gamma(B))$, we need to lift (\ref{eq:GenSol}) to the level of states, by virtue of (\ref{eq:expB}): 
\begin{equation}\label{eq:SolExp}\begin{split}
\omega\Big(\Lambda_t \big( \Gamma (B) \big)\Big) = \tr &\Big(e^{-(P - iH) t } B\, e^{- (P+iH) t }Q \Big)\\ &+ 2\, \int_0^t {\rm d}s\, \tr\Big(e^{-(P - iH) s } B\, e^{- (P+iH) s } A \Big)\, .
\end{split}
\end{equation}

An alternative perspective can be formulated by considering the object $\omega \circ \Lambda_t$; because $\Lambda_t$ describes a dynamical map, it actually follows that, for any $t>0$, $\omega \circ \Lambda_t$ is a quantum state in its own right. In other words, we can treat the dynamics in the Schr\"odinger picture, by defining a family of states
\begin{equation}\label{eq:omegat1}
\omega_t \equiv \omega \circ \Lambda_t\, .
\end{equation}
Intriguingly, equation (\ref{eq:SolExp}) even provides us with an explicit expression for the $Q(t)$ that appears in~(\ref{eq:expB}); by rewriting~(\ref{eq:SolExp}), we find
\begin{equation}\label{eq:omegat}
\begin{split}
&\omega_t \big(\Gamma(B)\big) = \tr \big(B Q(t)\big)\, ,\quad \text{with}  \\
&Q(t) = e^{- (P+iH) t }Q\,e^{-(P - iH) t } + 2\, \int_0^t {\rm d}s\, e^{- (P+iH) s } A\, e^{-(P - iH) s }\, ,
\end{split}
\end{equation}
where $Q$ is the single-particle covariance matrix for the initial state $\omega.$
 
Typically, at asymptotic times, pumped systems relax into a non-equilibrium steady state where finite currents are flowing.  This limiting state has completely forgotten the initial conditions of the system. Put differently, generically each system observable converges to a multiple of the identity. The way to describe the asymptotic state, is by explicitly considering the limit $t\rightarrow \infty$ in (\ref{eq:GenSol}). To do so, note that since $P >0$, generically,
\begin{equation}
\lim_{t \rightarrow \infty} e^{- (P+iH) t }Q\,e^{-(P - iH) t } = 0
\end{equation}
and therefore we find that
\begin{equation}\label{eq:defQNESSEtc}
\begin{split}
&\omega_{\rm NESS} \big(\Gamma(B)\big) = \tr \big(B Q_{\rm NESS}\big)\, ,\quad \text{with} \\
&Q_{\rm NESS} \equiv \lim_{t \rightarrow \infty} Q(t) = 2\, \int_0^{\infty} {\rm d}s\, e^{- (P+iH) s } A\, e^{-(P - iH) s }\, .
\end{split}
\end{equation}
Here NESS\ stands for non-equilibrium steady state. This,\eg implies that the expected number of particles in the system converges to
\begin{equation}
n = \omega_{\rm NESS}(N) = 2\, \int_0^{\infty} {\rm d}s\, \tr \Big(e^{- (P+iH) s } A\, e^{-(P - iH) s }\Big)\, .
\end{equation}
It is not hard to show \cite{davies_irreversible_1977} that the NESS state is the gauge-invariant quasi-free state determined by $Q_{\rm NESS}.$

\section{Currents}\label{sec:Currents}

Non-equilibrium systems are typically characterised by the presence of currents, even when they reach a stationary state. In this section, we first discuss general properties of currents, determined by the ``continuity equation'' (\ref{eq:balance}). We translate these results to a quantum mechanical setting to arrive at a sound definition of {\em quantum particle currents}. Finally, we extensively discuss a fundamental property of fermionic currents, which is one of our key results: {\em The existence of a universal upper bound}---irrespective of the specific potential encountered by the particle flow.

\subsection{Particle Currents}

We start by a formal definition of the particle current in the context of quantum master equations. The general structure in~(\ref{eq:Lindblad1}) presents us with the change of particles in the system over time. Because the number operator $N = \Gamma(\mathbb{1})$, we insert $B = \mathbb{1}$ into (\ref{eq:dyn1}) and obtain
\begin{align}
\frac{\rm d}{{\rm d}t} N &= \mathcal{D}^d(N) + \mathcal{D}^a(N)\, ,\label{thiseq1}\\
&\underrel{(\ref{eq:dyn1})}{=} -2\, \Gamma(D) - 2\, \Gamma(A) +2\, \tr \big(A\big) \1.\label{thiseq2}
\end{align}
Note that the Hamiltonian contribution vanishes in the evaluation of (\ref{eq:dyn1}) because $[\Gamma(H),\Gamma(\mathbb{1})] = \Gamma\big([H, \mathbb{1}]\big) = 0.$ This implies that the Hamiltonian, which is itself an observable of the form (\ref{quant:sec}), conserves the total number of particles.

We now study the particle current as a thermodynamic flux~\cite{alicki_detailed_1976,alicki_quantum_1979} and focus on its behaviour in the NESS~(\ref{eq:defQNESSEtc}). We note that, by definition of the {\em steady} state, the time derivative of the number operator is zero in the NESS, which yields, after combination of (\ref{thiseq1}, \ref{thiseq2}) with (\ref{eq:expB}), the balance equation
\begin{equation}\label{eq:balance}
\omega_{\rm NESS}\Big(\mathcal{D}^d(N) + \mathcal{D}^a(N)\Big) \underrel{(\ref{eq:defQNESSEtc})}{=} -2\,\tr \big(DQ_{\rm NESS}\big) +2\, \tr \big( A (\mathbb{1} - Q_{\rm NESS}) \big) = 0\, .
\end{equation}
We can now define the current of out flowing particles as 
\begin{equation}\label{eq:CurrentDefFin}
J \equiv \abs{\omega_{\rm NESS} \big( \mathcal{D}^d(N) \big)} = 2\,\tr \big(DQ_{\rm NESS}\big) \underrel{(\ref{eq:balance})}{=} 2\, \tr \big( A (\mathbb{1} - Q_{\rm NESS}) \big)\, ,
\end{equation}
where the absolute value is added because we focus on the magnitude of the current.

\subsection{Bounding the current}\label{sec:ProofBound}
 
Although expression~(\ref{eq:CurrentDefFin}) suggests that the current is independent of the Hamiltonian $H \in \mathcal{B}(\H)$, it is implicitly present in $Q_{\rm NESS}.$ Indeed, we can rewrite the current, using~(\ref{eq:defQNESSEtc}), to obtain
\begin{equation}\label{eq:FormalJ}
J = 4  \int^{\infty}_0 {\rm d}s  ~ \tr \Big( D e^{(-iH - P)s}  A e^{(iH - P)s}\big)\, .
\end{equation}
In principle, this expression allows for a direct computation of the current, although this is generically intricate, \eg when the operators in (\ref{eq:FormalJ}) do not commute. It is therefore instructive to derive a bound, to gain some general understanding of the parameter dependence of the current.

To do so, we first introduce the super-operator
\begin{equation}\label{eq:proof1}
\mathcal G(X) \equiv -i [H,X] + \{P,X\}
\enskip\text{with}\enskip
P = A + D\, .
\end{equation}
$\mathcal G$ can be split into a sum of two commuting terms, left multiplication by $P-iH$, and right multiplication by $P+iH$, respectively. Therefore, we may write
\begin{equation}
\mathcal G = \mathcal L_{P-iH} + \mathcal R_{P+iH}
\end{equation}
where
\begin{equation}
\mathcal L_Y(X) \equiv YX
\enskip\text{and}\enskip
\mathcal R_Y(X) \equiv XY\, .
\end{equation}
Generically, $\mathcal G$ is invertible and for positive definite $P>0$ we can use the identity
\begin{align}
\mathcal G^{-1}(X) 
&= \int_0^\infty \!\!\!{\rm d}s \exp(-s\mathcal G)(X) \nonumber \\
&= \int_0^\infty \!\!\!{\rm d}s \exp(-s\mathcal L_{P-iH}) \circ \exp(-s\mathcal R_{P+iH})(X) \nonumber \\ 
&= \int_0^\infty \!\!\!{\rm d}s\, e^{-s(P-iH)} X \, e^{-s(P+iH)}\, .
\end{align}
Next, we compute
\begin{equation}
\mathcal G( X^\dagger X) - \mathcal G(X^\dagger) X - X^\dagger \mathcal G(X) = -2 X^\dagger P X \le 0\, ,
\end{equation}
from which it follows that
\begin{equation}
\mathcal G( X^\dagger X) \le \mathcal G(X^\dagger) X + X^\dagger \mathcal G(X)\, .
\label{ineq}
\end{equation}
We now introduce a symmetrised zero temperature Duhamel (or Bogoliubov) inner product~\cite{fannes_correlation_1977,fannes_correlation_1977-1, petz_bogoliubov_1993,bratteli_operator_1997}:
\begin{equation}
\langle X, Y \rangle_\sim \equiv \tr \Bigl( X^\dagger 
\mathcal G^{-1}(Y) + \mathcal G^{-1}(X^\dagger) Y \Bigr)\, .
\end{equation}
Here $X$ and $Y$ are general linear operators. Positivity of the scalar product follows from the invertibility of $\mathcal G$, from $\mathcal G(X^\dagger) = \bigl(\mathcal G(X)\bigr)^\dagger$, from~(\ref{ineq}) and from
\begin{equation}
\tr\, \mathcal G(X^\dagger X) = \tr \{P,X^\dagger X\} \ge 0\, .
\end{equation}
For an explicit evaluation of Schwarz' inequality
\begin{equation}\label{eq:Schwarz}
\left| \langle A,P \rangle_\sim \right|^2 \leqslant \langle A,A\rangle_\sim\, \langle P,P\rangle_\sim
\end{equation}
we observe that
\begin{equation}
\mathcal L_{P \pm i H}(\1) + \mathcal R_{P \mp i H}(\1) = 2P\, .
\end{equation}
from which we infer that
\begin{align}\label{eq:proofFinal}
&\langle A,P\rangle_\sim =  \tr A, \\
&\langle P,P\rangle_\sim =  \tr P, \\
&\langle A,A\rangle_\sim = 2 \int_0^\infty \!\!\!{\rm d}s\, \tr \Bigl( A\, e^{-s(P-iH)}\, A\, e^{-s(P + iH)} \Bigr)\, \label{eq:proofFinalLast}.
\end{align}
Inserting these results in~(\ref{eq:Schwarz}), we obtain
\begin{equation}\label{eq:proofDiff}
\bigl(\tr (A) \bigr)^2 \leqslant \bigl( \tr (A) - J/2 \bigr) \tr (A + D)
\end{equation}
and it then follows that
\begin{equation}\label{bnd:1}
  \boxed{ J \leqslant 2 \frac{\tr (A)\, \tr (D)}{\tr (A + D)} =: J_{\max}\, ,}
\end{equation}
which is the desired bound to the current. It is a universal one, since it does only depend on the reservoir coupling agents $A$ and $D$, but not on the potential landscape which the fermions have to be transmitted through, defined by the system Hamiltonian $H$.

Since  $J_{\max}$ lacks a dependence on the single-particle Hamiltonian $H$, it is suggestive to inspect the tightness of the bound (\ref{bnd:1}) for variable relative strength of unitary dynamics and reservoir couplings.  For this purpose, we slightly rewrite (\ref{eq:FormalJ}) as 
\begin{equation}\label{eq:FormalJ2}
J = 4  \int^{\infty}_0 {\rm d}s  ~ \tr \Big( D\, e^{(-i\lambda H - P)s}  A\, e^{(i\lambda H - P)s}\big)\, ,
\end{equation}
where we introduced the parameter $\lambda$, to scale the relative strength of the Hamiltonian part of the dynamics as compared to particle loss and pump. $\lambda \to 0$ completely cancels the coherent part of the dynamics while $\lambda \to \infty$ makes the oscillatory Hamiltonian part much faster than the rates with which the systems couples to the reservoirs. 

\begin{figure}[t]
\centering
\includegraphics[width=0.8\textwidth]{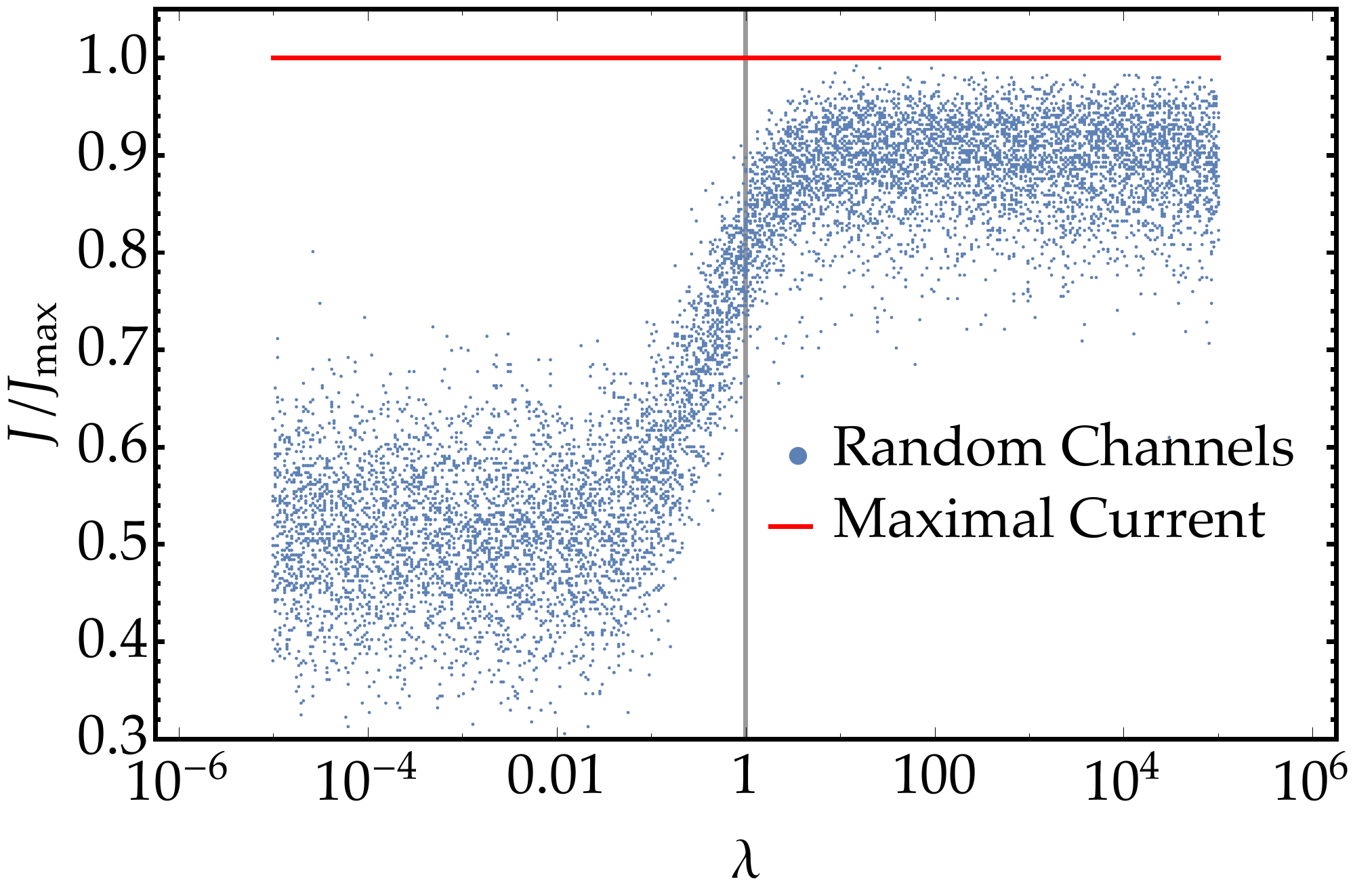}
\caption{\label{fig:Currents} Scatter plot of the stationary current $J$~(\ref{eq:FormalJ2}) relative to the maximal current $J_{\max}$~(\ref{bnd:1}). The variable $\lambda$~(\ref{eq:FormalJ2}) controls the relative strength of the Hamiltonian and incoherent contributions. For each data point $\log_{10} \lambda$ is randomly chosen from the interval $[-5,5]$. For each realisation, the Hamiltonian $H$ in~(\ref{eq:FormalJ2}) is chosen from the GOE~(\ref{eq:GOE}) with typical coupling $v/\sqrt{m}$ between modes, with $v = 1$, and mode number $m = 10$. The channels~(\ref{eq:dyn1}) $A$ and $D$ in~(\ref{eq:FormalJ2}) are drawn from a Wishart ensemble~(\ref{eq:Wishart}) with $m_A = 5$ and $m_D = 10$. Data points are compared to the upper bound $J=J_{\max}$ (horizontal red line). The value $\lambda = 1$ is indicated by a vertical grey line,  it coincides with the mean eigenvalue of $P=A+D$.}
\end{figure}

In the remainder of this section, we seek to numerically confirm the validity of~(\ref{bnd:1}) when $\lambda$ in~(\ref{eq:FormalJ2}) is varied. To approach this problem, we consider many realisations of the system, each time choosing a random $\lambda$, random Hamiltonian $H$, and random channels $A$ and $D$. For every realisation, the NESS current~(\ref{eq:FormalJ2}) is evaluated and compared to the upper bound~(\ref{bnd:1}). The results of this procedure are shown in Fig.~\ref{fig:Currents}, where specific choices for the random matrix ensembles were made: We consider a system of $m$ modes, \ie $\dim(\H) = m$. The Hamiltonian $H$ is sampled from the Gaussian orthogonal ensemble (GOE)~\cite{mehta_random_2004} which implies that it is a matrix whose entries are sampled from a normal distribution:
\begin{equation}\label{eq:GOE}
H_{ij} \sim {\rm Normal}\Big(0, (1+\delta_{ij}) \frac{v}{\sqrt{m}} \Big)\, .
\end{equation}
The parameter $v$ is related to the spectral radius (\ie~the largest eigenvalue in absolute value) and physically $v/\sqrt{m}$ can be thought of as the typical (\ie root mean squared) coupling strength between different modes. The matrices which describe the couplings between the system and the channels must be constructed so that they are always positive semi-definite. A standard method to generate random matrices fulfilling this constraint is to resort to the Wishart ensemble~\cite{wishart_generalised_1928}. The latter ensemble is solely determined by the number of absorption (dissipation) channels, $m_A$ ($m_D$). For our numerical simulations, we set 
\begin{equation}\label{eq:Wishart}
A = \frac{1}{m_A+m_D} W^{\dag}_aW_a, \quad\text{and} \quad D = \frac{1}{m_A+m_D}W^{\dag}_dW_d
\end{equation} 
where $W_a$ and $W_d$ are $m_A \times m$ and $m_D \times m$ matrices respectively. They are generated by choosing random components according to
\begin{equation}\label{eq:StandardNormWish}
\big(W_{a,d}\big)_{ij} \sim {\rm Normal}(0,1)\, .
\end{equation} 
The additional factor $(m_A+m_D)^{-1}$ in~(\ref{eq:Wishart}) is included to set the average eigenvalue of $P$~(\ref{eq:dyn1}) equal to $1$. With this choice of ensembles, we can genuinely interpret $\lambda$~(\ref{eq:FormalJ2}) as the ratio of the frequencies of the coherent oscillations induced by $H$ and the incoherent rates contained in $P.$ 

Fig.~\ref{fig:Currents} clearly shows that the bound~(\ref{bnd:1}) is valid for all realisations regardless of the magnitude $\lambda$. Nevertheless, we do observe that the bound is typically more accurate in the limit of dominantly coherent dynamics, characterised by $\lambda \gg 1$. In this contribution, we will not attempt to understand the specific statistical properties which are obtained from the random matrix theory treatment. We do, however, note that in the limit $\lambda \gg 1$ (or mathematically $\lambda \rightarrow \infty$), the current is susceptible to changes in the Hamiltonian (recall eq.~(\ref{eq:FormalJ2})) and that, therefore, a natural next step is to attempt to saturate the bound~(\ref{bnd:1}) in this regime.\footnote{From~(\ref{eq:FormalJ2}) it directly follows that attempts to saturate the bound by optimising $H$ are futile in the limit where $\lambda = 0.$}

\section{Symmetry enhanced current}\label{sec:Symmetry}

In this section, we investigate how an appropriate design of the system can generate a current close to $J_{\max}$~(\ref{bnd:1}). Because we are considering a designed system, it is reasonable to focus on the regime $\lambda \gg 1$ where the coherent dynamics has a strong influence on the current. To get a maximal effect of the coherent contributions, we rigorously focus on the regime $\lambda\to \infty$. This allows us to treat the problem using perturbation theory. In this limit, rapidly oscillating terms appear in~(\ref{eq:FormalJ2}) and by the Riemann-Lebesgue lemma~\cite{pedersen_analysis_1989} many contributions to $J$ cancel. 

The Hamiltonian can be represented in its spectral decomposition as
\begin{equation}
H = \sum_k E_k R_k\, .
\end{equation}
Here the $E_k$ are the eigenvalues of $H$ and $R_k$ are the orthogonal projectors on the corresponding eigenspaces of $H$. Using first order perturbation theory (where $1/\lambda$ is small), we compute
\begin{equation}\label{eq:Jpert}
\begin{split}
\lim_{\lambda \to \infty }  J
&= \lim_{\lambda \to \infty } 4 \int_0^{\infty} {\rm d}s ~\tr\Big\{D e^{-(P+i\lambda H) s}Ae^{- (P-i\lambda H) s} \Big\} \\
&= \sum_k 4\int_0^{\infty} {\rm d}s\, \tr\Big\{D R_k e^{-sR_kPR_k} A e^{-sR_kPR_k} R_k\Big\}\, .
\end{split}
\end{equation}
If we want to saturate the bound on the current we have to design the $R_k$ in an appropriate way.

Structuring Hamiltonians goes hand in hand with introducing symmetries. We therefore assume the existence of a unitary operator $U$, such that
\begin{equation}\label{eq:symHier}
[H,U] = 0\, .
\end{equation}
In order for such a symmetry to be useful, it must connect the couplings of the absorption channels $A$ to those of the output channels $D$, leading to the requirement
\begin{equation}\label{eq:balanceChannels}
U^{\dag}A\,U = D\, .
\end{equation}
Given these additional structures, we can rewrite~(\ref{eq:Jpert}) as
\begin{equation}
\lim_{\lambda \to \infty }  J = \sum_k 4\int_0^{\infty} {\rm d}s\, \tr\Big\{e^{-sR_kPR_k} R_k U^{\dag}A\, U\, R_k e^{-sR_kPR_k} A \Big\}\, .
\end{equation}
The fact that $U$ and $H$ commute, implies that $U$ is block-diagonal with respect to the spectral decomposition of $H$ 
\begin{equation}
U = \bigoplus_{k} U_k\, .
\end{equation} 
This further implies
\begin{equation}
\lim_{\lambda \to \infty }  J = \sum_k 4\int_0^{\infty} {\rm d}s\, \tr\Big\{e^{-sR_kPR_k} U^{\dag}_k R_k A\, R_k \, U_k e^{-sR_kPR_k} A \Big\}\, ,
\end{equation}
which can in general not be cast in a more transparent form. However, in the case where the Hamiltonian $H$ is non-degenerate (implying that, apart from (\ref{eq:symHier}), there are no unitary symmetries present in the system), we obtain that $R_k $ are {\em rank-one operators}. In this case we can express $U$ as
\begin{equation}
U = \sum_k e^{i \theta_k} R_k\, ,
\end{equation}
such that $e^{i \theta_k}$ are the eigenvalues of $U$. In turn this leads to
\begin{equation}
\lim_{\lambda \to \infty }  J = \sum_k 4\int_0^{\infty} {\rm d}s\, \tr\Big\{e^{-sR_kPR_k} R_k A\, R_k \, e^{-sR_kPR_k} A \Big\}\, .
\end{equation}
By virtue of (\ref{eq:Jpert}), where $D$ is replaced by $A$, and (\ref{eq:defQNESSEtc}), the right-hand side is exactly $\lim_{\lambda \to \infty} 2 \tr AQ_{\rm NESS}$. Due to the symmetry (\ref{eq:symHier}), this implies that  
\begin{equation}\label{eq:64}
\tr AQ_{\rm NESS} = \tr DQ_{\rm NESS} \quad \text{in the regime} \quad \lambda \to \infty\, .
\end{equation}
However, from the balance equation~(\ref{eq:balance}), we read that
\begin{equation}\label{eq:65}
\tr DQ_{\rm NESS} = \tr A -  \tr AQ_{\rm NESS}\, .
\end{equation}
Both equations (\ref{eq:64}) and (\ref{eq:65}) can hold simultaneously only when $ \tr AQ_{\rm NESS} =  \tr A/2,$ which implies that
\begin{equation}\label{eq:66}
\lim_{\lambda\rightarrow \infty} J = \tr A = \tr D\, .
\end{equation}
The second equality in (\ref{eq:66}) follows from~(\ref{eq:balanceChannels}). Inserting expression~(\ref{eq:balanceChannels}) into~(\ref{bnd:1}) directly yields
\begin{equation}
J_{\max} = \tr A \implies \lim_{\lambda\rightarrow \infty}J = J_{\max}
\end{equation}
which is exactly what we wanted to achieve.

In words, we have shown that, in the absence of degeneracies in $H$, it suffices to find a unitary operator $U$ which commutes with the Hamiltonian (\ie~a symmetry) and transforms $D$ into $A,$ in order to saturate the upper bound for the current in the limit $\lambda \to \infty.$ The most natural picture to associate with such a mathematical formulation, is that of a reflection symmetry. The limiting regime of $\lambda$ can be seen as a rigorous way of demanding weak coupling, implying that the time scales of the system dynamics are much faster than those set by the rates with which the system couples to its reservoirs. 

\begin{figure}[t]
\centering
\includegraphics[width=0.7\textwidth]{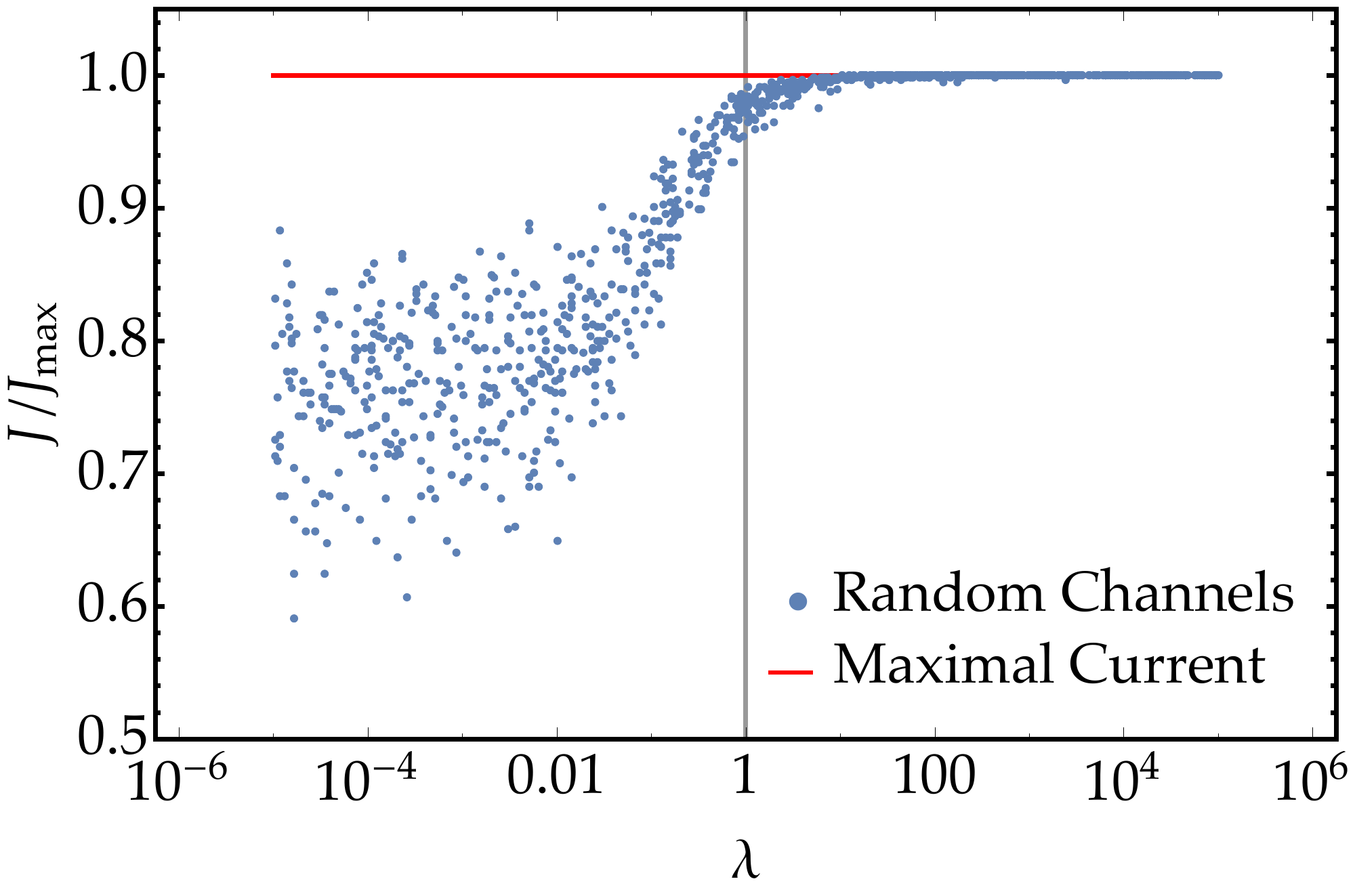}
\caption{\label{fig:CurrentsDesign} Scatter plot of the stationary current $J$~(\ref{eq:FormalJ2}) relative to the maximal current $J_{\max}$~(\ref{bnd:1}). The variable $\lambda$~(\ref{eq:FormalJ2}) controls the relative strength of the Hamiltonian and incoherent contributions. For each data point $\log_{10} \lambda$ is randomly chosen from the interval $[-5,5]$. Each realisation of the Hamiltonian $H$ in~(\ref{eq:FormalJ2}) is generated according to~(\ref{eq:generateHFromU}) with mode number $m = 10$ and random symmetry operator $U$ from the Haar measure~\cite{mehta_random_2004}. For each realisation, the absorption operator $A$~(\ref{eq:dyn1}) in~(\ref{eq:FormalJ2}) is drawn from a Wishart ensemble~(\ref{eq:Wishart}) with $m_A = 10$. The dissipation operator $D$ is determined by the condition~(\ref{eq:balanceChannels}). Data points are compared to the upper bound $J=J_{\max}$ (horizontal red line). The value $\lambda = 1$, which coincides with the mean eigenvalue of $P$, is indicated by a vertical grey line.}
\end{figure}

In realistic set-ups, this limit is never exactly achieved, therefore it is instructive to conduct numerical simulations to assess the deviations from the optimally achievable current, as a function of $\lambda$. This is done in Fig.~\ref{fig:CurrentsDesign}: To make the simulation as general as possible, we start by sampling the unitary $U$ introduced in (\ref{eq:symHier}) from the Haar measure~\cite{mehta_random_2004}.\footnote{The Haar measure may be interpreted as the uniform distribution over the set of unitary matrices.} Because the matrix is unitary and random, we can always obtain a spectral decomposition
\begin{equation}
U = \sum_{k=1}^m e^{i\theta_k} \ket{e_k}\bra{e_k}\, .
\end{equation}
We use this decomposition as the starting point for the construction of the Hamiltonian, which we define as
\begin{equation}\label{eq:generateHFromU}
H = \sum_{k=1}^m E_k \ket{e_k}\bra{e_k} \qquad E_k \sim {\rm Uniform}\big([-m/2, m/2]\big)\, .
\end{equation}
So $[H,U]=0$ follows by construction. The choice of the uniform distribution for the eigenvalues $E_k$ is arbitrary, it simply serves to ensure that the typical level spacing---and hence the typical frequency of the coherent oscillations---is independent of the system size $N$. We sample $A$ from the Wishart ensemble, see~(\ref{eq:Wishart}) and~(\ref{eq:StandardNormWish}), but must take the constraint $D = U^{\dag}AU$ into account. When focusing on the regime of $\lambda \ll 1$ in Fig.~\ref{fig:CurrentsDesign}, we observe a similar trend as for Fig.~\ref{fig:Currents}. However, once we approach $\lambda \gg 1,$ we observe that, indeed, $J \approx J_{\max}$ for all realisations.

The fact that the bound~(\ref{bnd:1}) can be saturated in the regime of dominantly coherent dynamics, $\lambda \gg 1,$ can be understood in a straightforward way: On the one hand, the rates with which the reservoirs couple to the system set time scales for particle exchange, which also governs the bound~(\ref{bnd:1}). On the other hand,  however, coherent time scales, set by the Hamiltonian, determine how the particles explore the various modes inside the system. Therefore, if these coherent time scales are too slow, particles will linger in the modes where they entered the system where they block the path for additional particles due to Pauli's exclusion principle. Hence, the limit $\lambda \gg 1,$ guarantees fast redistribution of particles within the system, and, in addition, the design principle (\ref{eq:balanceChannels}) and (\ref{eq:symHier}) guarantees a balance between input and output channels, such that particles can be extracted efficiently. In general, however, interference effects, incorporated in the fact that $A, D,$ and $H$ do not commute, make this naive picture more complicated. This is precisely why a mathematically rigorous treatment is important and non-trivial.

\begin{figure}[t]
\centering
\includegraphics[width=0.6\textwidth]{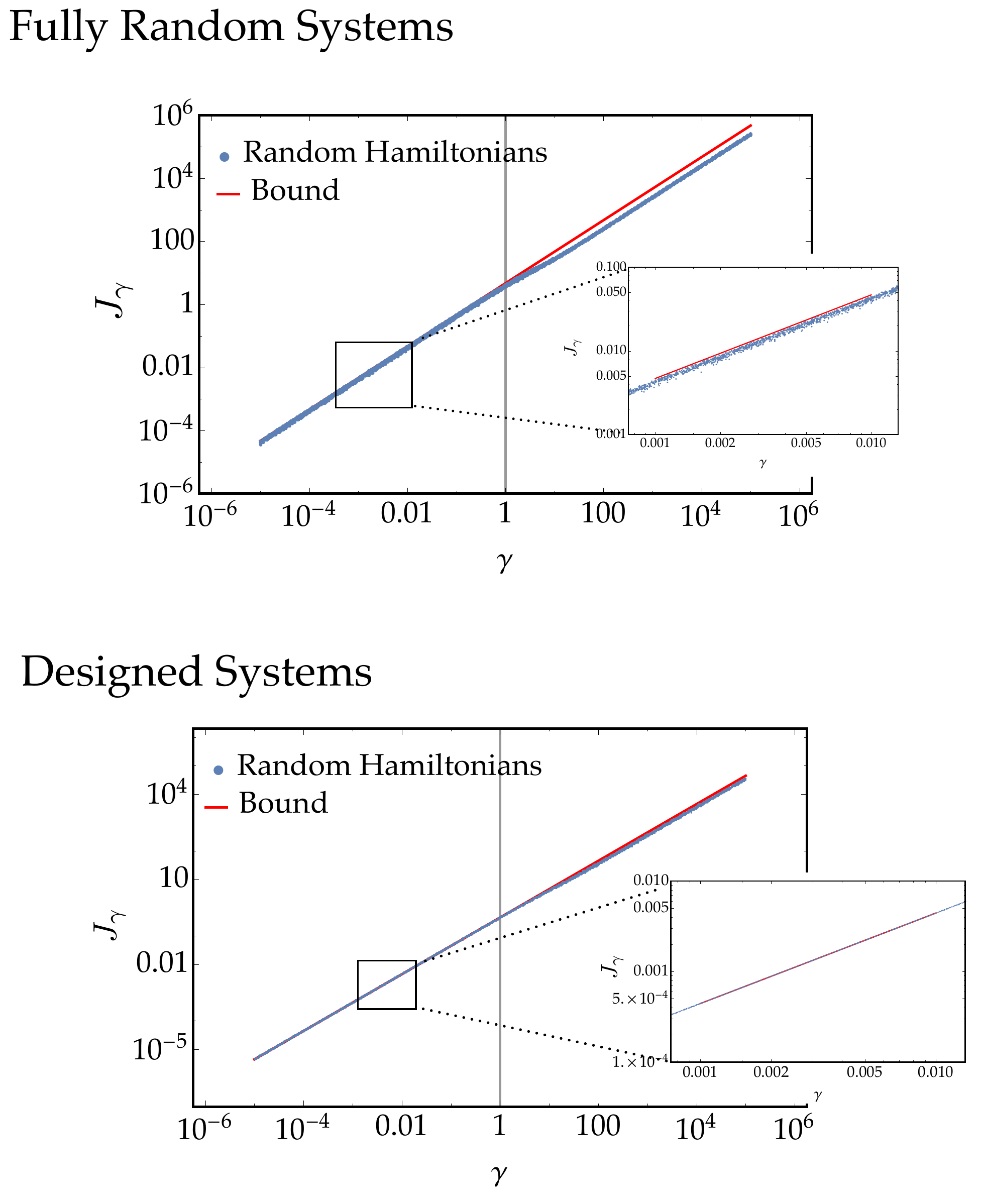}
\caption{\label{fig:CurrentsAbsolute} Scatter plot of the stationary current $J_{\gamma}$~(\ref{Jgamma}) in units of the mean-level spacing of the Hamiltonian. The variable $\gamma$~(\ref{Jgamma}) determines the incoherent time scales. For each data point $\log_{10} \gamma$ is randomly chosen from the interval $[-5,5]$. Data points are compared to the upper bound $J_{\max}$ (red line, \ref{bnd:1}). Both fully random and designed systems are shown. In each case, an inset shows the fluctuations of the currents $J_{\gamma}$ as compared to the bound, by zooming in on the parameter range $\gamma \in [0.001, 0.01]$. Hamiltonians and channels are generated as follows:\\
{\bf Fully random systems (top):} For each realisation, the Hamiltonian $H$ in~(\ref{Jgamma}) is chosen from the GOE~(\ref{eq:GOE}) with typical coupling $v/\sqrt{m}$ between modes, with $v = 1$, and mode number $m = 10$. A single set of channels~(\ref{eq:dyn1}) $A$ and $D$ in~(\ref{Jgamma}) are drawn from a Wishart ensemble~(\ref{eq:Wishart}) with $m_A = 5$ and $m_D = 10$, and remain fixed for all realisations.\\
{\bf Designed systems (bottom):} Each realisation of the Hamiltonian $H$ in~(\ref{Jgamma}) is generated according to~(\ref{eq:generateHFromU}) with mode number $m = 10$ and random symmetry operator $U$ from the Haar measure~\cite{mehta_random_2004}. A single absorption operator $A$~(\ref{eq:dyn1}) in~(\ref{eq:FormalJ2}) is drawn from a Wishart ensemble~(\ref{eq:Wishart}) with $m_A = 10$, which is kept fixed for all realisations. The dissipation operator $D$ is determined by the condition~(\ref{eq:balanceChannels}).}
\end{figure}

Furthermore, we stress that this discussion makes statements on the current $J$~(\ref{eq:FormalJ2}) relative to the bound $J_{\max}$~(\ref{bnd:1}). However, the maximal current $J_{\max}$ itself depends on the absorption and dissipation channels (as governed by operators $A$ and $D$). Therefore, when we rescale these parameters as $A \rightarrow \gamma A$ and $D \rightarrow \gamma D$, it directly follows, from expression~(\ref{bnd:1}), that $J_{\max}\rightarrow \gamma J_{\max}$. The results of Fig.~\ref{fig:CurrentsDesign} imply that for any such value of $\gamma$, the current $J$~(\ref{eq:FormalJ2}) can be optimised, for $\lambda \gg \gamma$, by appropriately designing the system according to (\ref{eq:balanceChannels}) and (\ref{eq:symHier}). However, because the value of the bound increases with $\gamma$, it is conceivable that a large value of $\gamma$ (and thus large rates of particle exchange between system and reservoirs) can lead to large currents, even for slow coherent time scales, \ie $\lambda<\gamma$. This hypothesis is, indeed, confirmed in Fig.~\ref{fig:CurrentsAbsolute}. We can define the current $J_{\gamma}$ which results from rescaling $A \rightarrow \gamma A$ and $D \rightarrow \gamma D$:
\begin{equation}\label{Jgamma}
J_{\gamma} = 4  \int^{\infty}_0 {\rm d}s  ~ \tr \Big( \gamma D\, e^{(-i H - \gamma P)s}  \gamma A\, e^{(i H - \gamma P)s}\big)\, .
\end{equation}
Fig.~\ref{fig:CurrentsAbsolute} shows how the current (in units which are fixed by the Hamiltonian's mean-level spacing) scales as a function of the rescaling parameter $\gamma$. The Hamiltonians are generated following (\ref{eq:GOE}) and (\ref{eq:generateHFromU}), for fully random and designed Hamiltonians, respectively. For the fully random systems, a single set of input and output channels, represented by $A$ and $D$, respectively, is randomly chosen according to (\ref{eq:Wishart}) and kept fixed. In the simulation of the designed systems we fix only $A$ and generate $D$ according to (\ref{eq:balanceChannels}).

In Fig.~\ref{fig:CurrentsAbsolute} we see that typically the current increases when we increase the incoherent rates for particle exchange (by varying $\gamma$). However, we also observe that the bound is tighter in the regime of dominantly coherent dynamics as given by $\gamma \ll 1$. In this regime, the designed systems give rise to optimal transport by saturating the bound, whereas we see fluctuations in the full random systems (see inset in Fig.~\ref{fig:CurrentsAbsolute}). It must be noted that the double logarithmic scale of the plot masks these fluctuation.

Up to this point, we studied systems which contain a finite number of particles at all times. The strength of the C*-algebraic treatment and the formulation of the model in terms of a general CAR is the possibility to extend the setting to systems with an infinite number of degrees of freedom. In the following section, we consider systems that require a technical treatment based on current densities. We prove a generalisation of the bound~(\ref{bnd:1}) to a class of shift-invariant systems as commonly encountered in theoretical solid-state physics.

\section{The Quantum Ribbon}\label{sec:Ribbon}

Above we focused on systems with a finite dimensional mode space, which excludes models with a translational invariance in some spatial directions. The latter situation requires to perform a thermodynamic limit, \ie we first have to consider a finite subsystem, and subsequently perform a limiting procedure where the size of the system tends to infinity while the particle density remains finite \cite{bratteli_operator_1997}. We now consider such a model situation, with some inspiration from~\cite{mosonyi_asymptotic_2008}.

The specific system under consideration is a ribbon: A 2D system with translation invariance in one direction, and finite width in the orthogonal dimension. We assume that the system is accurately described by a tight-binding model and therefore the single-particle Hilbert space is given by a discrete lattice $\H := l^2(\mathbb{Z}) \otimes \mathbb{C}^d,$ where $d$ quantifies the finite width of the lattice. For $k \in \mathbb{Z}$ we denote by $\mathbf{1}_{\{k\}}$ the sequence in $l^2(\mathbb{Z})$ with 1 at place $k$ and 0 everywhere else. The mode space of our system can then also be seen as $\bigoplus_{k \in \mathbb{Z}} \mathbf{1}_{\{k\}} \otimes \mathbb{C}^d$ with the one-step shift along the ribbon given by $\mathbf{1}_{\{k\}} \otimes \psi \mapsto \mathbf{1}_{\{k + 1\}} \otimes \psi.$

\subsection{Shift-Invariance}

Let us first focus on the space $l^2(\mathbb{Z})$. The {\em Fourier transform} $F: l^2(\mathbb{Z}) \rightarrow L^2\big([0,2\pi)\big)$ can be defined through its action on the indicator functions $\mathbf{1}_{\{k\}} \in l^2(\mathbb{Z})$:
\begin{equation}
F \mathbf{1}_{\{k\}} := \phi_k, \quad \text{with} \quad \phi_k(x) = e^{ikx}\, .
\end{equation}
A bounded operator operator $A$ on $l^2(\mathbb{Z})$ is shift-invariant if and only if $A = F^{-1} M_{\hat{a}} F$. Here
\begin{equation}
(M_{\hat{a}}\psi)(x) = \hat{a}(x)\, \psi(x)\, ,\quad \psi \in L^2\big([0,2\pi)\big)
\end{equation}
and $\hat{a} \in L^\infty\big([0,2\pi)\big).$ Therefore a bounded shift-invariant operator on $l^2(\mathbb{Z})$ corresponds to a multiplication operator on $L^2\big([0,2\pi)\big)$ by a bounded function on $[0,2\pi)$. Hermitian operators correspond hereby to real-valued functions and positive semi-definite operators to non-negative functions.

This can straightforwardly be generalised to $\H =  l^2(\mathbb{Z}) \otimes \mathbb{C}^d$: We say that a bounded operator $X$ on $l^2(\mathbb{Z}) \otimes \mathbb{C}^d$ is translation-invariant along the ribbon iff $X= ( F^{-1} \otimes \mathbb{1}) M_{\hat{X}} (F \otimes \mathbb{1})$, where $\hat{X}: [0,2\pi) \rightarrow \mathcal{M}_d$ is a bounded matrix-valued function. If we now denote $e_{k,l} = \mathbf{1}_{\{k\}}\otimes e_l,$ with $\{e_l\}$ the standard basis in $\mathbb{C}^d$, we may write that
\begin{equation}\label{eq:shiftInv1}
\braket{e_{k',l'}}{X\,e_{k,l}} = \frac{1}{2 \pi} \int^{2\pi}_0 {\rm d}x \braket{e_{l'}}{\hat{X}(x) \, e_l} e^{-i(k'-k) x}\, .
\end{equation} 
It also follows that, for two shift-invariant operators $X$ and $Y,$ 
\begin{equation}\label{eq:shiftandshift}
XY = (F^{-1} \otimes \mathbb{1} ) M_{\hat{X}\hat{Y}} (F \otimes \mathbb{1})\, .
\end{equation}
It is useful to generalise~(\ref{eq:shiftInv1}) to the case where $\hat{X}: [0,2\pi) \rightarrow \mathcal{M}_d$ is integrable.\footnote{The integrability of a matrix-valued function can be understood in a component-wise way.} In general, such a choice leads to an unbounded $X$.   

To discuss currents, we are confronted with the problem that the global number operator $N$, which counts the number of fermions on the ribbon, is not an element in the CAR algebra over $l^2(\mathbb{Z}) \otimes \mathbb{C}^d$. In fact, $\mathcal{A}$ does not contain any
shift-invariant elements except for the multiples of $\mathbb{1}.$ Shift-invariant elements are introduced by their local restrictions on 
finite subsets $\Lambda \subset \mathbb{Z}$ of the ribbon. To construct these local restriction, we define the appropriate projectors 
\begin{equation}\label{eq:Pn}
P_\Lambda := \sum_{p \in \Lambda} \ket{\mathbf{1}_{\{p\}}}\bra{\mathbf{1}_{\{p\}}} \otimes \mathbb{1}\, .
\end{equation}
We can now consider the $\Lambda$-restriction $\Gamma(P_\Lambda X P_\Lambda)$ of ``$\Gamma(X)$'', which is a \emph{bona fide} element of the algebra. Translation-invariance manifests itself by $\Gamma(P_{\Lambda + 1} X P_{\Lambda + 1})$ being the one-step shift of $\Gamma(P_\Lambda X P_\Lambda)$. The global number operator corresponds to the choice $\hat X(x) = 1$ for $x \in [0,2\pi)$ and its restriction to $\Lambda$ is just the number operator for the mode space $l^2(\Lambda) \otimes \mathbb{C}^d,$ \ie it counts the number of fermions on the compact domain defined by the restriction $\Lambda$.

Suppose that a shift-invariant $0 \leqslant Q \leqslant \mathbb{1}$ determines the one-particle expectations (\ref{eq:Q}) and that $X$ defines a shift-invariant one-particle observable as above. Both $Q$ and $X$ are determined by matrix-valued functions $\hat Q$ and $\hat X$ on $[0,2\pi)$ that satisfy the requirements that $0 \leqslant \hat Q \leqslant 1$ and $\hat X$ be real-valued and integrable. We can now consider the expectation of the density of $\Gamma(X)$ in a specific state $\omega_{Q}$ where we rewrite henceforth $P_n \equiv P_{\{1,\ldots, n\}}$
\begin{equation}
\tilde{x}(\omega_{Q}) = \lim_{n \to \infty} \frac{1}{n}\, \omega_Q\bigl( \Gamma(P_n X P_n) \bigr)\, , 
\end{equation}
where we introduce the ``$\sim$'' to refer to densities in the system. Because of translation-invariance, there is no problem in fixing the leftmost site of the interval at 1. A small computation, similar to the type of computations used in proving Szeg\"o's theorem \cite{mosonyi_asymptotic_2008}, yields
\begin{equation}\label{eq:szeg}
\tilde{x}(\omega_Q) = \frac{1}{2\pi}\, \int_0^{2\pi} {\rm d}x \tr_{\mathbb{C}^d}\, \bigl( \hat X(x)\, \hat Q(x) \bigr) \, .
\end{equation}      
  
\subsection{Currents in the Quantum Ribbon}

The bound~(\ref{bnd:1}) on the fermionic current nicely fits with shift-invariance. For shift-invariant $H$, $A$, and $D,$ both sides of the bound scale linearly with the length of the sub-interval on the ribbon that we consider. It then suffices to renormalise the inequality to obtain an analogous bound for densities.  

The dynamics is {\em a priori} similar to the dynamics generated by~(\ref{eq:dyn1}), although we now specifically focus on the situation where $H, A,$ and $D$ are shift-invariant operators. We can simply repeat the arguments from Section \ref{sec:NESS} above and obtain that in the long-time limit any state asymptotically converges to a shift-invariant state determined by the matrix-valued function
\begin{equation}
x \mapsto \hat Q_{\rm NESS}(x) = 2\, \int_0^{\infty} {\rm d}s\, e^{- (\hat P(x) + i \hat H(x)) s } \hat A(x)\, e^{-(\hat P(x) - i \hat H(x)) s }\, .
\end{equation}  

The particle density $\rho(\omega)$ for a translation-invariant state determined by $\hat Q$ (\ref{eq:szeg}) is given by
\begin{equation}
\tilde{\rho}(\omega) = \frac{1}{2\pi} \int^{2\pi}_0 {\rm d}x \, \frac{1}{d} \tr_{\mathbb{C}^d}\big\{\hat{Q}(x)\big\}\, .
\end{equation} 
Note that in our dynamical system we are typically dealing with a particle density $\tilde{\rho}(\omega)$ that changes over time.

We start by considering the evolution of the local number operator $N_n = \Gamma(P_n)$, which is described by (\ref{eq:dyn1}):
\begin{equation}
\frac{\rm d}{{\rm d} t} \Gamma(P_n) = i \Gamma\Big([H, P_n] - \{A+D, P_n \}\Big) +2\, \tr \big(P_n A P_n\big) \1
\end{equation}
such that 
\begin{equation}
\begin{split}
\frac{\rm d}{{\rm d} t} \omega\big(\Gamma(P_n)\big) &= i \tr ([H, P_n]Q) - \tr(\{A+D, P_n\}Q) +2\, \tr \big(P_n A P_n\big) \\
&= \tr (P_n [Q,H] P_n) - \tr(P_n\{Q,A+D\}P_n) + 2\, \tr \big(P_n A P_n\big) \, .
\end{split}
\end{equation}
Note that, because $P_n$ is a projector, (\ref{eq:expB}) yields $\omega\big(\Gamma(P_n)\big) = \tr \big(P_n Q P_n\big)$, and that~(\ref{eq:shiftandshift}) implies that commutators and anti-commutators of shift-invariant operators are again shift-invariant. Therefore, we may use~(\ref{eq:szeg}) and evaluate
\begin{equation}
\begin{split}
\frac{\rm d}{{\rm d} t} \tilde{\rho}(\omega) 
&= \lim_{n \rightarrow \infty} \frac{1}{n\,d} \frac{\rm d}{{\rm d} t}\tr \big(P_n Q P_n\big) \\
&= \frac{1}{2 \pi \, d}\int_0^{2\pi} {\rm d} x\, 2\Big(\tr_{\mathbb{C}^d}\big\{\hat{A}(x)(\mathbb{1}-\hat{Q}(x))\big\}-\tr_{\mathbb{C}^d} \big\{\hat{D}(x)\hat{Q}(x)\big\}\Big)
\end{split}
\end{equation}
where we already used that $\tr_{\mathbb{C}^d}(i [\hat{Q}(x),\hat{H}(x)]) = 0$ for all $x.$

By definition of the non-equilibrium steady state
\begin{equation}
\frac{\rm d}{{\rm d} t} \tilde{\rho}(\omega_{\rm NESS}) = 0 
\end{equation}
and we therefore define the current density as 
\begin{equation}
\tilde{j} = \frac{1}{\pi \, d}\int_0^{2\pi} {\rm d} x\, \tr_{\mathbb{C}^d} \big\{\hat{D}(x)\hat{Q}(x)\big\} = \frac{1}{\pi \, d}\int_0^{2\pi} {\rm d} x\, \tr_{\mathbb{C}^d}\big\{\hat{A}(x)(\mathbb{1}-\hat{Q}(x))\big\}\, .
\end{equation}
For every $x \in [0,2\pi),$ we can apply~(\ref{eq:Schwarz}) to find
\begin{equation}\label{eq:ineqHere}
\bigl(\tr_{\mathbb{C}^d} \hat{A}(x) \bigr)^2 \leqslant \Bigl( \tr_{\mathbb{C}^d} \hat{A}(x) - \tr_{\mathbb{C}^d}\big\{\hat{A}(x)(\mathbb{1}-\hat{Q}(x))\big\} \Bigr) \tr_{\mathbb{C}^d} \big\{\hat{A}(x) + \hat{D}(x)\big\}
\end{equation}
which can be rewritten and integrated to obtain
\begin{equation}
\boxed{\tilde{j} \leqslant \frac{1}{\pi \, d}\int_0^{2\pi} {\rm d} x\,\frac{\tr_{\mathbb{C}^d} \hat{A}(x)\tr_{\mathbb{C}^d} \hat{D}(x)}{\tr_{\mathbb{C}^d} \hat{A}(x) + \tr_{\mathbb{C}^d} \hat{D}(x)}\, ,}
\end{equation}
as a universal upper bound for the fermionic current across the quantum ribbon.

\section{Bosonic systems}\label{sec:Bosons}

Throughout the preceding parts of this contribution, we focused on systems of non-interacting fermions. Our methods are, however, also applicable to systems of non-interacting bosons. In this scenario, we must consider additional technical details related to the algebra of canonical commutation relations (CCR)~\cite{petz_invitation_1990,bratteli_operator_1997,verbeure_many-body_2011}. One technical issue is that, for infinite dimensional mode spaces $\H$, the bosonic algebra only allows us to define creation and annihilation operators in a representation dependent way. Another technical issue is that, even for finite dimensional $\H$, states are not necessarily given by a density matrix on Fock space. Therefore, we here deliberately focus on systems with a finite number of particles, such that we remain in the Fock representation at all times. 

The bosonic Fock space is defined (quite analogous to (\ref{eq:FockFerm1}, \ref{eq:FockFerm3})) as
\begin{equation} 
\Gamma^b(\H) := \mathbb{C} \oplus \H \oplus \H^{(2)} \oplus \H^{(3)} \oplus \cdots
\end{equation}
with 
\begin{equation}
\H^{(n)} = \H \otimes \H \otimes \cdots \otimes \H \Big\vert_{\rm sym} \, .
\end{equation}
The CCR can now we written in terms of non-local creation and annihilation operators $b^\dagger(\phi)$ and $b(\phi)$ which act on ``Slater permanents'' in a similar fashion as~(\ref{cre:CAR}) and~(\ref{anh:CAR}). These unbounded operators on $\Gamma^b(\H)$ satisfy canonical commutation relations:
\begin{align}
&\psi \in \H \mapsto b^{\dag}(\psi) \quad \text{is $\mathbb{C}$-linear} \\
&[b(\phi), b^{\dag}(\psi)] = \braket{\phi}{\psi}\1 \qquad \text{and} \qquad [b(\phi),b(\psi)] = 0\, .
\end{align} 

In analogy to the fermionic case, we describe our dynamics in terms of the phenomenological master equation\footnote{We define this equation for all linear operators on Fock space, a set which we denote ${\rm Lin}\big(\Gamma(\H)\big)$.}
\begin{equation}\label{eq:LindbladFormDynamicsBosons}
\begin{split}
&\frac{\rm d}{{\rm d} t} X = \mathcal{D}(X) = -i[\Gamma(H),X] + \mathcal{D}^a(X) + \mathcal{D}^d(X),  \quad\text{ for all } X \in {\rm Lin}\big(\Gamma(\H)\big)\, ,\\
&\text{with} \quad \mathcal{D}^{a/d}(X)  =  \sum_{i} {L^{a/d}_i}^{\dag} X L^{a/d}_i - \frac{1}{2} \{{L^{a/d}_i}^{\dag} L^{a/d}_i , X\}\, ,
\end{split}
\end{equation}
and, in analogy to~(\ref{eq:Lindblad1}-\ref{eq:Lindblad3}) we choose $L^d_i = \sqrt{\gamma^d_i} b(\delta_i)$ and $L^a_i = \sqrt{\gamma^a_i} b^{\dag}(\alpha_i).$ Again, $\{\alpha_i \in \H \}$ denote the single-particle state vectors in which particles are absorbed into the system, whereas $\{\delta_i \in \H\}$ denote the single-particle state vectors from which particles are dissipated out of the system. Because the creation and annihilation operators are unbounded, it remains to verify that this leads to a valid dynamical map, \ie that it fulfils the conditions (\ref{eq:dynMapCond}) and maps elements of the algebra onto other elements of the alegbra. To do so, we evaluate
\begin{equation}\begin{split}\label{eq:LindbladRewrittenBosons}
\mathcal{D}(b^{\dag}(\phi)b(\psi)) = b^{\dag}\big((iH - P) \phi\big)b\big(\psi\big) + b^{\dag}\big(\phi\big)b\big((iH-P)\psi\big) + 2 \braket{\phi}{A \psi} \1,
\end{split}
\end{equation} 
where the bosonic $P$ is defined as 
\begin{equation}
\label{eq:Pbosons}
P\equiv D-A\, ,
\end{equation}
with $A$ and $D$ as in~(\ref{eq:dyn1}).
A fundamental difference between the fermionic and bosonic case is that the bosonic $P$ in~(\ref{eq:Pbosons}) is not necessarily a positive semi-definite operator on the single-particle space. 

We use~(\ref{eq:LindbladRewrittenBosons}) to evaluate the dynamics of a general single-particle observable
\begin{equation}
\Gamma^b(B) = \sum_{i,j} \braket{\eta_i}{B \eta_j} b^{\dag}(\eta_i)b(\eta_j)\, ,
\end{equation}
where $\{\eta_i\}$ forms an orthonormal basis of the single-particle Hilbert space $\H.$ Straightforward integration of (\ref{eq:LindbladFormDynamicsBosons}) leads to
\begin{equation}\label{eq:dynamicsSinglePartBoson}
\Lambda_t(\Gamma^b (B)) = \Gamma^b\left( e^{t(iH - P)} B e^{t(-iH - P)} \right)+ \tr\left( \int^t_0 {\rm d}s  ~ 2 A e^{(iH - P)s} B e^{(-iH - P)s} \right)\1\, .
\end{equation}
We now observe that the case where $P$ is {\em not} positive semi-definite can lead to severe problems because it typically does not allow for the system to remain contained within the Fock representation at all times. This can be understood by assessing the time evolution of the particle number expectation value. We consider systems which are initially local with respect to the Fock representation, therefore the state is given by a density matrix $\rho$ which acts on $\Gamma^b(\H)$ and
\begin{equation}
\< N\>_{\rho} = \tr \{\rho \, \Gamma^b(\mathbb{1})\} < \infty\, .
\end{equation}
However, when $P$ is not positive semi-definite, for generic $\rho$ the asymptotic particle number is given by 
\begin{equation}
\lim_{t\to\infty} \tr \{\rho \, \Lambda_t(\Gamma^b(\mathbb{1}))\} = \infty\, .
\end{equation}
and thus diverges in the long-time limit. Physically this means that the system is unstable and never reaches a steady state. Therefore, we must impose that 
\begin{equation}\label{eq:bosonicStability}
D \geqslant A\, ,
\end{equation}
and therefore $P \geqslant 0$, to ensure that the system remain confined to Fock space for all  all times. This implies that systems of non-interacting bosons which absorb particles from an external reservoir require a sufficient (as quantified by~(\ref{eq:bosonicStability})) amount of dissipation to ensure the existence of a well-defined NESS.\footnote{What is quite obvious from a physical point of view, think, \eg of a micro-maser scenario \cite{lugiato_connection_1987}  where a single quantised mode of the radiation field is pumped by a sequence of two-level atoms which enter the resonator in the excited state and interact resonantly with the resonator mode. The steady state is there defined by an equilibrium of gain and loss (damping) of the resonator mode. Note, however, that in this paradigmatic realisation the effective pump rate depends nonlinearly on the occupation number of the resonator mode.}

Having imposed condition~(\ref{eq:bosonicStability}), we find that the solutions to the bosonic and fermionic equations of motion, (\ref{eq:dynamicsSinglePartBoson}) and (\ref{eq:GenSol}), respectively, are very similar, such that the same analysis as above can be repeated. The bosonic continuity equation is the same as the fermionic one when we write it in terms of $P$: In the NESS we find
\begin{equation}\label{eq:outflowingChannels}
2 \tr PQ_{\rm NESS} =  2 \tr A\, .
\end{equation}
However, the definition of $P$ has changed, so that the following balance equation between incoming and outflowing currents holds:
\begin{equation}\label{eq:currentsFinal}
2 \tr\, \Big(A  (\mathbb{1} + Q_{\rm NESS})\Big) = 2 \tr \, (DQ_{\rm NESS})\, .
\end{equation}
This implies that we can still describe the current flowing through the system as $J \equiv 2 \tr \, (DQ_{\rm NESS}).$ 

Remarkably, this definition of the current, together with~(\ref{eq:bosonicStability}), implies that we can next precisely follow the steps (\ref{eq:proof1}-\ref{eq:proofFinalLast}) of the proof for the fermionic bound. However, in (\ref{eq:proofDiff}) we did employ the explicit form of the continuity equation, and therefore this step differs from the present bosonic case. We now find that
\begin{equation}
(\tr A)^2 \leqslant  (J/2 - \tr A) \tr P\, ,
\end{equation}
which implies
\begin{equation}
\tr A \,(\tr A + \tr P) \leqslant \frac{J}{2}\, \tr P
\end{equation}
and therefore
\begin{equation}\label{eq:Jmin}
\boxed{J \geqslant 2\frac{\tr A \tr D}{\tr (D-A)} =: J_{\rm min}\, .}
\end{equation}
The inequality~(\ref{eq:Jmin}) is remarkable because its derivation is largely analogous to that for the fermionic case, but it ultimately produces a very different phenomenology: There is no upper bound for bosonic currents in the NESS. However, bosonic currents are always stronger than a given quantity $J_{\min}$ which is set by the channels. In systems where $A$ comes close to $D$, while respecting (\ref{eq:bosonicStability}), we see that the rate at which particles stream through the system can become arbitrarily large. 

\begin{figure}[t]
\centering
\includegraphics[width=0.69\textwidth]{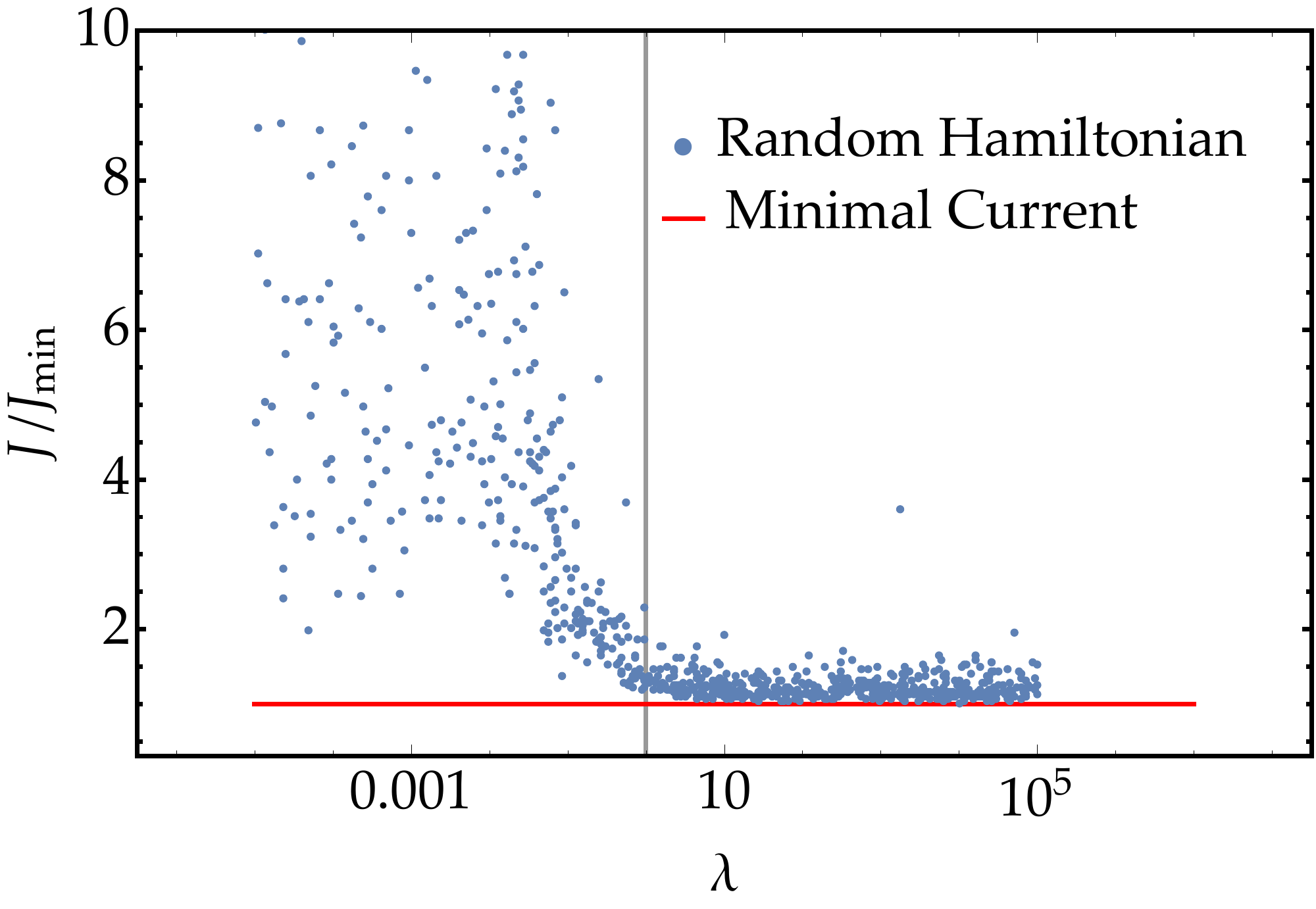}
\caption{\label{fig:CurrentBosons}Scatter plot which benchmarks the stationary current $J$~(\ref{eq:JlambdaBosons}) with respect to the minimal current $J_{\min}$~(\ref{eq:Jmin}). The variable $\lambda$~(\ref{eq:JlambdaBosons}) controls the relative strength of the Hamiltonian with respect to incoherent contributions; for each data point, $\log_{10} \Lambda$ is randomly chosen from the interval $[-5,5].$ The Hamiltonians $H$ in~(\ref{eq:JlambdaBosons}) are chosen from the GOE~(\ref{eq:GOE}) with typical interaction $v/\sqrt{m}$ between modes, with $v = 1$, and mode number $m = 10$. The channels $P$~(\ref{eq:Pbosons}) and $A$~(\ref{eq:dyn1}) in~(\ref{eq:JlambdaBosons}) are drawn from a Wishart ensemble~(\ref{eq:Wishart}) with $m_A = 5$ and $m_P = 10$. $D$ in~(\ref{eq:JlambdaBosons}) is directly obtained from~(\ref{eq:Pbosons}). Data points are compared to the lower bound $J=J_{\min}$ (horizontal red line). The value $\lambda = 1$ is indicated (vertical grey line) since it represents the typical incoherent rate as the mean eigenvalue of $P$.}
\end{figure}

Finally, we numerically scrutinize the lower bound~(\ref{eq:Jmin}). These results shown in Fig.~\ref{fig:CurrentBosons} are obtained through evaluation of the exact expression for $J$:
\begin{equation}\label{eq:JlambdaBosons}
J = 4 \int^{\infty}_0 {\rm d}s \, \tr \Big\{ e^{(-i\lambda H - P)s}A e^{(i\lambda H - P)s} D\Big\}\, ,
\end{equation}
where $P$ is given by~(\ref{eq:Pbosons}). The parameter $\lambda$ serves the same purpose as in~(\ref{eq:FormalJ2}) and Figs.~\ref{fig:Currents} and \ref{fig:CurrentsDesign}, and the simulations are performed in a similar fashion as for Fig.~\ref{fig:Currents}: The value $\lambda$ is chosen randomly in a way such that $\log_{10} \lambda$ is uniformly distributed, whereas the Hamiltonians are sampled from the GOE~(\ref{eq:GOE}). The choice of $A$ and $D$ is more subtle because of condition~(\ref{eq:bosonicStability}). To satisfy this constraint, we rather choose $P$ and $A$ from the Wishart ensemble~(\ref{eq:Wishart}) to subsequently determine $D = P + A.$ 

The results in Fig.~\ref{fig:CurrentBosons} confirm the prediction by the lower bound~(\ref{eq:Jmin}) and show a drastically different behaviour compared to the fermionic case of Fig.~\ref{fig:Currents}. These results can be understood as a manifestation of quantum statistics. However, bosons do not disturb each other statistically when they start piling up (as happens when  $\tr (D-A) \approx 0$). Where the fermionic ``repulsion'' is often more important than the particle interaction, this is not the case for bosons. Hence, the assumption of absence of interactions for bosons is rather more stringent than it is for fermions. Therefore, one should be careful when interpreting these bosonic results when $\tr (D-A)$ is small and particle densities become high.

\section{Conclusions}

We described many-fermion and many-boson systems in which particles are incoherently pumped in and dissipated from the system, such that the total dynamics can be considered to be Markovian (memoryless). We prove that, in the absence of interactions between particles, the total particle current across the system exhibits universal properties in the stationary state: We could derive an upper bound~(\ref{bnd:1}) for fermionic currents, and, under some additional conditions which prevent the system from unlimited heating, a lower bound~(\ref{eq:Jmin}) for bosonic currents. Remarkably, both bounds are independent of the specific potential landscape the particles are transmitted through.

Numerically, we showed that, though counterintuitive, the bounds are typically sharp in the regime where the coherent dynamics' frequencies are high compared to the incoherent rates which determine the time scales of the reservoir coupling. This also led us to design Hamiltonians, as generators of the coherent dynamics, which saturate the bound in the limit where coherent dynamics is dominant. We proved that, in this limit, very general symmetry properties imposed onto the Hamiltonian suffice to achieve our goal. More specifically, we considered a unitary operator that commutes with the Hamiltonian and maps channels which connect the incoming reservoirs to the system onto channels which connect the system to the outgoing reservoirs. With these {\em design principles}, we can saturate our upper bound for {\em fermionic currents}. We note that the centro-symmetry~\cite{christandl_perfect_2005, zech_centrosymmetry_2014,walschaers_statistical_2015,ortega_efficient_2016}, discussed in the context of optimal transport, is a special case of our present design principle. Hence, this work also improves the understanding of how such symmetries enhance quantum transport. 

Our results offer a starting point for the investigation of several new questions, ranging from the relation of the here presented results to the Landauer formalism~\cite{imry_introduction_2009,landauer_spatial_1957} to applications, \eg in the quantum transport theory of disordered systems \cite{beenakker_random-matrix_1997,schlawin_bunching_2012,kropf_effective_2016} or in the quantum statistics of non-equilibrium dynamical processes~\cite{uzdin_equivalence_2015,uzdin_universal_2014}. On a more fundamental level, the natural next steps are to investigate \cite{walschaers_efficient_2016} how particle-interactions or other general sources of dephasing~\cite{mendoza-arenas_dephasing_2013,kordas_non-equilibrium_2015} impact the here derived universal bounds. In addition, it is a natural question to wonder what happens when the assumption of Markovian dynamics breaks down, \eg it was recently shown  \cite{wang_nonequilibrium_2015}, for a non-equilibrium spin-boson model, that the current is optimal for an intermediate coupling between system and reservoirs.

\section*{Acknowledgements}
M.W. thanks Joris Van Houtven and Oliver Lauwers for useful discussions during their project on ``Kwantumtransport in open kwantumnetwerken''.  M.W. Acknowledges financial support by the German National Academic Foundation. 

\bibliography{Paper_Currents}

\begin{thebibliography}{100}%
\makeatletter
\providecommand \@ifxundefined [1]{%
 \@ifx{#1\undefined}
}%
\providecommand \@ifnum [1]{%
 \ifnum #1\expandafter \@firstoftwo
 \else \expandafter \@secondoftwo
 \fi
}%
\providecommand \@ifx [1]{%
 \ifx #1\expandafter \@firstoftwo
 \else \expandafter \@secondoftwo
 \fi
}%
\providecommand \natexlab [1]{#1}%
\providecommand \enquote  [1]{``#1''}%
\providecommand \bibnamefont  [1]{#1}%
\providecommand \bibfnamefont [1]{#1}%
\providecommand \citenamefont [1]{#1}%
\providecommand \href@noop [0]{\@secondoftwo}%
\providecommand \href [0]{\begingroup \@sanitize@url \@href}%
\providecommand \@href[1]{\@@startlink{#1}\@@href}%
\providecommand \@@href[1]{\endgroup#1\@@endlink}%
\providecommand \@sanitize@url [0]{\catcode `\\12\catcode `\$12\catcode
  `\&12\catcode `\#12\catcode `\^12\catcode `\_12\catcode `\%12\relax}%
\providecommand \@@startlink[1]{}%
\providecommand \@@endlink[0]{}%
\providecommand \url  [0]{\begingroup\@sanitize@url \@url }%
\providecommand \@url [1]{\endgroup\@href {#1}{\urlprefix }}%
\providecommand \urlprefix  [0]{URL }%
\providecommand \Eprint [0]{\href }%
\providecommand \doibase [0]{http://dx.doi.org/}%
\providecommand \selectlanguage [0]{\@gobble}%
\providecommand \bibinfo  [0]{\@secondoftwo}%
\providecommand \bibfield  [0]{\@secondoftwo}%
\providecommand \translation [1]{[#1]}%
\providecommand \BibitemOpen [0]{}%
\providecommand \bibitemStop [0]{}%
\providecommand \bibitemNoStop [0]{.\EOS\space}%
\providecommand \EOS [0]{\spacefactor3000\relax}%
\providecommand \BibitemShut  [1]{\csname bibitem#1\endcsname}%
\let\auto@bib@innerbib\@empty
\bibitem [{\citenamefont {Steinlesberger}\ \emph {et~al.}(2002)\citenamefont
  {Steinlesberger}, \citenamefont {Engelhardt}, \citenamefont {Schindler},
  \citenamefont {Steinh{\"o}gl}, \citenamefont {von Glasow}, \citenamefont
  {Mosig},\ and\ \citenamefont {Bertagnolli}}]{steinlesberger_electrical_2002}%
  \BibitemOpen
  \bibfield  {author} {\bibinfo {author} {\bibfnamefont {G.}~\bibnamefont
  {Steinlesberger}}, \bibinfo {author} {\bibfnamefont {M.}~\bibnamefont
  {Engelhardt}}, \bibinfo {author} {\bibfnamefont {G.}~\bibnamefont
  {Schindler}}, \bibinfo {author} {\bibfnamefont {W.}~\bibnamefont
  {Steinh{\"o}gl}}, \bibinfo {author} {\bibfnamefont {A.}~\bibnamefont {von
  Glasow}}, \bibinfo {author} {\bibfnamefont {K.}~\bibnamefont {Mosig}}, \ and\
  \bibinfo {author} {\bibfnamefont {E.}~\bibnamefont {Bertagnolli}},\ }\href
  {\doibase 10.1016/S0167-9317(02)00815-8} {\bibfield  {journal} {\bibinfo
  {journal} {Microelectronic Engineering}\ }\bibinfo {series} {{MAM}2002},\
  \textbf {\bibinfo {volume} {64}},\ \bibinfo {pages} {409} (\bibinfo {year}
  {2002})}\BibitemShut {NoStop}%
\bibitem [{\citenamefont {Imry}(2009)}]{imry_introduction_2009}%
  \BibitemOpen
  \bibfield  {author} {\bibinfo {author} {\bibfnamefont {Y.}~\bibnamefont
  {Imry}},\ }\href@noop {} {{\selectlanguage {English}\emph {\bibinfo {title}
  {Introduction to mesoscopic physics}}}}\ (\bibinfo  {publisher} {Oxford
  University Press},\ \bibinfo {address} {Oxford},\ \bibinfo {year}
  {2009})\BibitemShut {NoStop}%
\bibitem [{\citenamefont {Moors}\ \emph {et~al.}(2015)\citenamefont {Moors},
  \citenamefont {Sor{\'e}e},\ and\ \citenamefont
  {Magnus}}]{moors_modeling_2015}%
  \BibitemOpen
  \bibfield  {author} {\bibinfo {author} {\bibfnamefont {K.}~\bibnamefont
  {Moors}}, \bibinfo {author} {\bibfnamefont {B.}~\bibnamefont {Sor{\'e}e}}, \
  and\ \bibinfo {author} {\bibfnamefont {W.}~\bibnamefont {Magnus}},\ }\href
  {\doibase 10.1063/1.4931573} {\bibfield  {journal} {\bibinfo  {journal} {J.
  Appl. Phys.}\ }\textbf {\bibinfo {volume} {118}},\ \bibinfo {pages} {124307}
  (\bibinfo {year} {2015})}\BibitemShut {NoStop}%
\bibitem [{\citenamefont {Moors}\ \emph {et~al.}(2016)\citenamefont {Moors},
  \citenamefont {Sor{\'e}e},\ and\ \citenamefont
  {Magnus}}]{moors_resistivity_2016}%
  \BibitemOpen
  \bibfield  {author} {\bibinfo {author} {\bibfnamefont {K.}~\bibnamefont
  {Moors}}, \bibinfo {author} {\bibfnamefont {B.}~\bibnamefont {Sor{\'e}e}}, \
  and\ \bibinfo {author} {\bibfnamefont {W.}~\bibnamefont {Magnus}},\ }\href
  {http://arxiv.org/abs/1606.05972} {\bibfield  {journal} {\bibinfo  {journal}
  {arXiv:1606.05972 [cond-mat]}\ } (\bibinfo {year} {2016})},\ \bibinfo {note}
  {arXiv: 1606.05972}\BibitemShut {NoStop}%
\bibitem [{\citenamefont {Ashcroft}\ and\ \citenamefont
  {Mermin}(1976)}]{ashcroft_solid_1976}%
  \BibitemOpen
  \bibfield  {author} {\bibinfo {author} {\bibfnamefont {N.~W.}\ \bibnamefont
  {Ashcroft}}\ and\ \bibinfo {author} {\bibfnamefont {N.~D.}\ \bibnamefont
  {Mermin}},\ }\href@noop {} {\emph {\bibinfo {title} {Solid state physics}}}\
  (\bibinfo  {publisher} {Holt, Rinehart and Winston},\ \bibinfo {address} {New
  York},\ \bibinfo {year} {1976})\BibitemShut {NoStop}%
\bibitem [{\citenamefont {Derezi{\'n}ski}\ \emph {et~al.}(2008)\citenamefont
  {Derezi{\'n}ski}, \citenamefont {Roeck},\ and\ \citenamefont
  {Maes}}]{derezinski_fluctuations_2008}%
  \BibitemOpen
  \bibfield  {author} {\bibinfo {author} {\bibfnamefont {J.}~\bibnamefont
  {Derezi{\'n}ski}}, \bibinfo {author} {\bibfnamefont {W.~D.}\ \bibnamefont
  {Roeck}}, \ and\ \bibinfo {author} {\bibfnamefont {C.}~\bibnamefont {Maes}},\
  }\href {\doibase 10.1007/s10955-008-9500-8} {\bibfield  {journal} {\bibinfo
  {journal} {J. Stat. Phys.}\ }\textbf {\bibinfo {volume} {131}},\ \bibinfo
  {pages} {341} (\bibinfo {year} {2008})}\BibitemShut {NoStop}%
\bibitem [{\citenamefont {Asadian}\ \emph {et~al.}(2013)\citenamefont
  {Asadian}, \citenamefont {Manzano}, \citenamefont {Tiersch},\ and\
  \citenamefont {Briegel}}]{asadian_heat_2013}%
  \BibitemOpen
  \bibfield  {author} {\bibinfo {author} {\bibfnamefont {A.}~\bibnamefont
  {Asadian}}, \bibinfo {author} {\bibfnamefont {D.}~\bibnamefont {Manzano}},
  \bibinfo {author} {\bibfnamefont {M.}~\bibnamefont {Tiersch}}, \ and\
  \bibinfo {author} {\bibfnamefont {H.~J.}\ \bibnamefont {Briegel}},\ }\href
  {\doibase 10.1103/PhysRevE.87.012109} {\bibfield  {journal} {\bibinfo
  {journal} {Phys. Rev. E}\ }\textbf {\bibinfo {volume} {87}},\ \bibinfo
  {pages} {012109} (\bibinfo {year} {2013})}\BibitemShut {NoStop}%
\bibitem [{\citenamefont {Manzano}\ \emph {et~al.}(2012)\citenamefont
  {Manzano}, \citenamefont {Tiersch}, \citenamefont {Asadian},\ and\
  \citenamefont {Briegel}}]{manzano_quantum_2012}%
  \BibitemOpen
  \bibfield  {author} {\bibinfo {author} {\bibfnamefont {D.}~\bibnamefont
  {Manzano}}, \bibinfo {author} {\bibfnamefont {M.}~\bibnamefont {Tiersch}},
  \bibinfo {author} {\bibfnamefont {A.}~\bibnamefont {Asadian}}, \ and\
  \bibinfo {author} {\bibfnamefont {H.~J.}\ \bibnamefont {Briegel}},\ }\href
  {\doibase 10.1103/PhysRevE.86.061118} {\bibfield  {journal} {\bibinfo
  {journal} {Phys. Rev. E}\ }\textbf {\bibinfo {volume} {86}},\ \bibinfo
  {pages} {061118} (\bibinfo {year} {2012})}\BibitemShut {NoStop}%
\bibitem [{\citenamefont {Mendoza-Arenas}\ \emph {et~al.}(2013)\citenamefont
  {Mendoza-Arenas}, \citenamefont {Grujic}, \citenamefont {Jaksch},\ and\
  \citenamefont {Clark}}]{mendoza-arenas_dephasing_2013}%
  \BibitemOpen
  \bibfield  {author} {\bibinfo {author} {\bibfnamefont {J.~J.}\ \bibnamefont
  {Mendoza-Arenas}}, \bibinfo {author} {\bibfnamefont {T.}~\bibnamefont
  {Grujic}}, \bibinfo {author} {\bibfnamefont {D.}~\bibnamefont {Jaksch}}, \
  and\ \bibinfo {author} {\bibfnamefont {S.~R.}\ \bibnamefont {Clark}},\ }\href
  {\doibase 10.1103/PhysRevB.87.235130} {\bibfield  {journal} {\bibinfo
  {journal} {Phys. Rev. B}\ }\textbf {\bibinfo {volume} {87}},\ \bibinfo
  {pages} {235130} (\bibinfo {year} {2013})}\BibitemShut {NoStop}%
\bibitem [{\citenamefont {Kordas}\ \emph {et~al.}(2015)\citenamefont {Kordas},
  \citenamefont {Witthaut},\ and\ \citenamefont
  {Wimberger}}]{kordas_non-equilibrium_2015}%
  \BibitemOpen
  \bibfield  {author} {\bibinfo {author} {\bibfnamefont {G.}~\bibnamefont
  {Kordas}}, \bibinfo {author} {\bibfnamefont {D.}~\bibnamefont {Witthaut}}, \
  and\ \bibinfo {author} {\bibfnamefont {S.}~\bibnamefont {Wimberger}},\ }\href
  {\doibase 10.1002/andp.201400189} {\bibfield  {journal} {\bibinfo  {journal}
  {Ann. Phys. (Berlin)}\ }\textbf {\bibinfo {volume} {527}},\ \bibinfo {pages}
  {619} (\bibinfo {year} {2015})}\BibitemShut {NoStop}%
\bibitem [{\citenamefont {Thingna}\ \emph {et~al.}(2016)\citenamefont
  {Thingna}, \citenamefont {Manzano},\ and\ \citenamefont {Cao}}]{Ref11}%
  \BibitemOpen
  \bibfield  {author} {\bibinfo {author} {\bibfnamefont {J.}~\bibnamefont
  {Thingna}}, \bibinfo {author} {\bibfnamefont {D.}~\bibnamefont {Manzano}}, \
  and\ \bibinfo {author} {\bibfnamefont {J.}~\bibnamefont {Cao}},\ }\href
  {http://dx.doi.org/10.1038/srep28027} {\bibfield  {journal} {\bibinfo
  {journal} {Scientific Reports}\ }\textbf {\bibinfo {volume} {6}},\ \bibinfo
  {pages} {28027 EP } (\bibinfo {year} {2016})}\BibitemShut {NoStop}%
\bibitem [{\citenamefont {Manzano}\ \emph {et~al.}(2016)\citenamefont
  {Manzano}, \citenamefont {Chuang},\ and\ \citenamefont
  {Cao}}]{manzano_quantum_2016}%
  \BibitemOpen
  \bibfield  {author} {\bibinfo {author} {\bibfnamefont {D.}~\bibnamefont
  {Manzano}}, \bibinfo {author} {\bibfnamefont {C.}~\bibnamefont {Chuang}}, \
  and\ \bibinfo {author} {\bibfnamefont {J.}~\bibnamefont {Cao}},\ }\href
  {\doibase 10.1088/1367-2630/18/4/043044} {\bibfield  {journal} {\bibinfo
  {journal} {New J. Phys.}\ }\textbf {\bibinfo {volume} {18}},\ \bibinfo
  {pages} {043044} (\bibinfo {year} {2016})}\BibitemShut {NoStop}%
\bibitem [{\citenamefont {Engel}\ \emph {et~al.}(2007)\citenamefont {Engel},
  \citenamefont {Calhoun}, \citenamefont {Read}, \citenamefont {Ahn},
  \citenamefont {Man{\v c}al}, \citenamefont {Cheng}, \citenamefont
  {Blankenship},\ and\ \citenamefont {Fleming}}]{engel_evidence_2007}%
  \BibitemOpen
  \bibfield  {author} {\bibinfo {author} {\bibfnamefont {G.~S.}\ \bibnamefont
  {Engel}}, \bibinfo {author} {\bibfnamefont {T.~R.}\ \bibnamefont {Calhoun}},
  \bibinfo {author} {\bibfnamefont {E.~L.}\ \bibnamefont {Read}}, \bibinfo
  {author} {\bibfnamefont {T.-K.}\ \bibnamefont {Ahn}}, \bibinfo {author}
  {\bibfnamefont {T.}~\bibnamefont {Man{\v c}al}}, \bibinfo {author}
  {\bibfnamefont {Y.-C.}\ \bibnamefont {Cheng}}, \bibinfo {author}
  {\bibfnamefont {R.~E.}\ \bibnamefont {Blankenship}}, \ and\ \bibinfo {author}
  {\bibfnamefont {G.~R.}\ \bibnamefont {Fleming}},\ }\href {\doibase
  10.1038/nature05678} {\bibfield  {journal} {\bibinfo  {journal} {Nature}\
  }\textbf {\bibinfo {volume} {446}},\ \bibinfo {pages} {782} (\bibinfo {year}
  {2007})}\BibitemShut {NoStop}%
\bibitem [{\citenamefont {Cheng}\ and\ \citenamefont
  {Fleming}(2009)}]{cheng_dynamics_2009}%
  \BibitemOpen
  \bibfield  {author} {\bibinfo {author} {\bibfnamefont {Y.-C.}\ \bibnamefont
  {Cheng}}\ and\ \bibinfo {author} {\bibfnamefont {G.~R.}\ \bibnamefont
  {Fleming}},\ }\href {\doibase 10.1146/annurev.physchem.040808.090259}
  {\bibfield  {journal} {\bibinfo  {journal} {Annu. Rev. Phys. Chem.}\ }\textbf
  {\bibinfo {volume} {60}},\ \bibinfo {pages} {241} (\bibinfo {year}
  {2009})}\BibitemShut {NoStop}%
\bibitem [{\citenamefont {Scholes}\ \emph {et~al.}(2011)\citenamefont
  {Scholes}, \citenamefont {Fleming}, \citenamefont {Olaya-Castro},\ and\
  \citenamefont {van Grondelle}}]{scholes_lessons_2011}%
  \BibitemOpen
  \bibfield  {author} {\bibinfo {author} {\bibfnamefont {G.~D.}\ \bibnamefont
  {Scholes}}, \bibinfo {author} {\bibfnamefont {G.~R.}\ \bibnamefont
  {Fleming}}, \bibinfo {author} {\bibfnamefont {A.}~\bibnamefont
  {Olaya-Castro}}, \ and\ \bibinfo {author} {\bibfnamefont {R.}~\bibnamefont
  {van Grondelle}},\ }\href {\doibase 10.1038/nchem.1145} {\bibfield  {journal}
  {\bibinfo  {journal} {Nat Chem}\ }\textbf {\bibinfo {volume} {3}},\ \bibinfo
  {pages} {763} (\bibinfo {year} {2011})}\BibitemShut {NoStop}%
\bibitem [{\citenamefont {Scholak}\ \emph {et~al.}(2011)\citenamefont
  {Scholak}, \citenamefont {de~Melo}, \citenamefont {Wellens}, \citenamefont
  {Mintert},\ and\ \citenamefont {Buchleitner}}]{scholak_efficient_2011}%
  \BibitemOpen
  \bibfield  {author} {\bibinfo {author} {\bibfnamefont {T.}~\bibnamefont
  {Scholak}}, \bibinfo {author} {\bibfnamefont {F.}~\bibnamefont {de~Melo}},
  \bibinfo {author} {\bibfnamefont {T.}~\bibnamefont {Wellens}}, \bibinfo
  {author} {\bibfnamefont {F.}~\bibnamefont {Mintert}}, \ and\ \bibinfo
  {author} {\bibfnamefont {A.}~\bibnamefont {Buchleitner}},\ }\href {\doibase
  10.1103/PhysRevE.83.021912} {\bibfield  {journal} {\bibinfo  {journal} {Phys.
  Rev. E}\ }\textbf {\bibinfo {volume} {83}},\ \bibinfo {pages} {021912}
  (\bibinfo {year} {2011})}\BibitemShut {NoStop}%
\bibitem [{\citenamefont {Alicki}\ and\ \citenamefont
  {Miklaszewski}(2012)}]{alicki_resonance_2012}%
  \BibitemOpen
  \bibfield  {author} {\bibinfo {author} {\bibfnamefont {R.}~\bibnamefont
  {Alicki}}\ and\ \bibinfo {author} {\bibfnamefont {W.}~\bibnamefont
  {Miklaszewski}},\ }\href {\doibase 10.1063/1.3697975} {\bibfield  {journal}
  {\bibinfo  {journal} {J. Chem. Phys}\ }\textbf {\bibinfo {volume} {136}},\
  \bibinfo {pages} {134103} (\bibinfo {year} {2012})}\BibitemShut {NoStop}%
\bibitem [{\citenamefont {Walschaers}\ \emph {et~al.}(2013)\citenamefont
  {Walschaers}, \citenamefont {Diaz}, \citenamefont {Mulet},\ and\
  \citenamefont {Buchleitner}}]{walschaers_optimally_2013}%
  \BibitemOpen
  \bibfield  {author} {\bibinfo {author} {\bibfnamefont {M.}~\bibnamefont
  {Walschaers}}, \bibinfo {author} {\bibfnamefont {J.~F.-d.-C.}\ \bibnamefont
  {Diaz}}, \bibinfo {author} {\bibfnamefont {R.}~\bibnamefont {Mulet}}, \ and\
  \bibinfo {author} {\bibfnamefont {A.}~\bibnamefont {Buchleitner}},\ }\href
  {\doibase 10.1103/PhysRevLett.111.180601} {\bibfield  {journal} {\bibinfo
  {journal} {Phys. Rev. Lett.}\ }\textbf {\bibinfo {volume} {111}},\ \bibinfo
  {pages} {180601} (\bibinfo {year} {2013})}\BibitemShut {NoStop}%
\bibitem [{\citenamefont {Levi}\ \emph {et~al.}(2015)\citenamefont {Levi},
  \citenamefont {Mostarda}, \citenamefont {Rao},\ and\ \citenamefont
  {Mintert}}]{levi_quantum_2015}%
  \BibitemOpen
  \bibfield  {author} {\bibinfo {author} {\bibfnamefont {F.}~\bibnamefont
  {Levi}}, \bibinfo {author} {\bibfnamefont {S.}~\bibnamefont {Mostarda}},
  \bibinfo {author} {\bibfnamefont {F.}~\bibnamefont {Rao}}, \ and\ \bibinfo
  {author} {\bibfnamefont {F.}~\bibnamefont {Mintert}},\ }\href {\doibase
  10.1088/0034-4885/78/8/082001} {\bibfield  {journal} {\bibinfo  {journal}
  {Rep. Prog. Phys.}\ }\textbf {\bibinfo {volume} {78}},\ \bibinfo {pages}
  {082001} (\bibinfo {year} {2015})}\BibitemShut {NoStop}%
\bibitem [{\citenamefont {Blankenship}(2002)}]{blankenship_molecular_2002}%
  \BibitemOpen
  \bibfield  {author} {\bibinfo {author} {\bibfnamefont {R.~E.}\ \bibnamefont
  {Blankenship}},\ }\href@noop {} {\emph {\bibinfo {title} {Molecular
  mechanisms of photosynthesis}}}\ (\bibinfo  {publisher} {Blackwell Science},\
  \bibinfo {address} {Oxford ; Malden, MA},\ \bibinfo {year}
  {2002})\BibitemShut {NoStop}%
\bibitem [{\citenamefont {Jesenko}\ and\ \citenamefont {{\v Z}nidari{\v
  c}}(2012)}]{jesenko_optimal_2012}%
  \BibitemOpen
  \bibfield  {author} {\bibinfo {author} {\bibfnamefont {S.}~\bibnamefont
  {Jesenko}}\ and\ \bibinfo {author} {\bibfnamefont {M.}~\bibnamefont {{\v
  Z}nidari{\v c}}},\ }\href {\doibase 10.1088/1367-2630/14/9/093017} {\bibfield
   {journal} {\bibinfo  {journal} {New J. Phys.}\ }\textbf {\bibinfo {volume}
  {14}},\ \bibinfo {pages} {093017} (\bibinfo {year} {2012})}\BibitemShut
  {NoStop}%
\bibitem [{\citenamefont {Jesenko}\ and\ \citenamefont {{\v Z}nidari{\v
  c}}(2013)}]{jesenko_excitation_2013}%
  \BibitemOpen
  \bibfield  {author} {\bibinfo {author} {\bibfnamefont {S.}~\bibnamefont
  {Jesenko}}\ and\ \bibinfo {author} {\bibfnamefont {M.}~\bibnamefont {{\v
  Z}nidari{\v c}}},\ }\href {\doibase 10.1063/1.4802816} {\bibfield  {journal}
  {\bibinfo  {journal} {J. Chem. Phys}\ }\textbf {\bibinfo {volume} {138}},\
  \bibinfo {pages} {174103} (\bibinfo {year} {2013})}\BibitemShut {NoStop}%
\bibitem [{\citenamefont {Manzano}(2013)}]{manzano_quantum_2013}%
  \BibitemOpen
  \bibfield  {author} {\bibinfo {author} {\bibfnamefont {D.}~\bibnamefont
  {Manzano}},\ }\href {\doibase 10.1371/journal.pone.0057041} {\bibfield
  {journal} {\bibinfo  {journal} {PLoS ONE}\ }\textbf {\bibinfo {volume} {8}},\
  \bibinfo {pages} {e57041} (\bibinfo {year} {2013})}\BibitemShut {NoStop}%
\bibitem [{\citenamefont {Witt}\ and\ \citenamefont
  {Mintert}(2013)}]{witt_stationary_2013}%
  \BibitemOpen
  \bibfield  {author} {\bibinfo {author} {\bibfnamefont {B.}~\bibnamefont
  {Witt}}\ and\ \bibinfo {author} {\bibfnamefont {F.}~\bibnamefont {Mintert}},\
  }\href {\doibase 10.1088/1367-2630/15/9/093020} {\bibfield  {journal}
  {\bibinfo  {journal} {New J. Phys.}\ }\textbf {\bibinfo {volume} {15}},\
  \bibinfo {pages} {093020} (\bibinfo {year} {2013})}\BibitemShut {NoStop}%
\bibitem [{\citenamefont {Shatokhin}\ \emph {et~al.}(2016)\citenamefont
  {Shatokhin}, \citenamefont {Walschaers}, \citenamefont {Schlawin},\ and\
  \citenamefont {Buchleitner}}]{shatokhin_coherence_2016}%
  \BibitemOpen
  \bibfield  {author} {\bibinfo {author} {\bibfnamefont {V.~N.}\ \bibnamefont
  {Shatokhin}}, \bibinfo {author} {\bibfnamefont {M.}~\bibnamefont
  {Walschaers}}, \bibinfo {author} {\bibfnamefont {F.}~\bibnamefont
  {Schlawin}}, \ and\ \bibinfo {author} {\bibfnamefont {A.}~\bibnamefont
  {Buchleitner}},\ }\href {http://arxiv.org/abs/1602.07878} {\bibfield
  {journal} {\bibinfo  {journal} {arXiv:1602.07878 [quant-ph]}\ } (\bibinfo
  {year} {2016})},\ \bibinfo {note} {arXiv: 1602.07878}\BibitemShut {NoStop}%
\bibitem [{\citenamefont {Reed}\ \emph {et~al.}(1988)\citenamefont {Reed},
  \citenamefont {Randall}, \citenamefont {Aggarwal}, \citenamefont {Matyi},
  \citenamefont {Moore},\ and\ \citenamefont {Wetsel}}]{reed_observation_1988}%
  \BibitemOpen
  \bibfield  {author} {\bibinfo {author} {\bibfnamefont {M.~A.}\ \bibnamefont
  {Reed}}, \bibinfo {author} {\bibfnamefont {J.~N.}\ \bibnamefont {Randall}},
  \bibinfo {author} {\bibfnamefont {R.~J.}\ \bibnamefont {Aggarwal}}, \bibinfo
  {author} {\bibfnamefont {R.~J.}\ \bibnamefont {Matyi}}, \bibinfo {author}
  {\bibfnamefont {T.~M.}\ \bibnamefont {Moore}}, \ and\ \bibinfo {author}
  {\bibfnamefont {A.~E.}\ \bibnamefont {Wetsel}},\ }\href {\doibase
  10.1103/PhysRevLett.60.535} {\bibfield  {journal} {\bibinfo  {journal} {Phys.
  Rev. Lett.}\ }\textbf {\bibinfo {volume} {60}},\ \bibinfo {pages} {535}
  (\bibinfo {year} {1988})}\BibitemShut {NoStop}%
\bibitem [{\citenamefont {Beenakker}(1991)}]{beenakker_theory_1991}%
  \BibitemOpen
  \bibfield  {author} {\bibinfo {author} {\bibfnamefont {C.~W.~J.}\
  \bibnamefont {Beenakker}},\ }\href {\doibase 10.1103/PhysRevB.44.1646}
  {\bibfield  {journal} {\bibinfo  {journal} {Phys. Rev. B}\ }\textbf {\bibinfo
  {volume} {44}},\ \bibinfo {pages} {1646} (\bibinfo {year}
  {1991})}\BibitemShut {NoStop}%
\bibitem [{\citenamefont {Jalabert}\ \emph {et~al.}(1994)\citenamefont
  {Jalabert}, \citenamefont {Pichard},\ and\ \citenamefont
  {Beenakker}}]{jalabert_universal_1994}%
  \BibitemOpen
  \bibfield  {author} {\bibinfo {author} {\bibfnamefont {R.~A.}\ \bibnamefont
  {Jalabert}}, \bibinfo {author} {\bibfnamefont {J.-L.}\ \bibnamefont
  {Pichard}}, \ and\ \bibinfo {author} {\bibfnamefont {C.~W.~J.}\ \bibnamefont
  {Beenakker}},\ }\href {\doibase 10.1209/0295-5075/27/4/001} {\bibfield
  {journal} {\bibinfo  {journal} {EPL}\ }\textbf {\bibinfo {volume} {27}},\
  \bibinfo {pages} {255} (\bibinfo {year} {1994})}\BibitemShut {NoStop}%
\bibitem [{\citenamefont {Contreras-Pulido}\ \emph {et~al.}(2012)\citenamefont
  {Contreras-Pulido}, \citenamefont {Splettstoesser}, \citenamefont
  {Governale}, \citenamefont {K{\"o}nig},\ and\ \citenamefont
  {B{\"u}ttiker}}]{contreras-pulido_time_2012}%
  \BibitemOpen
  \bibfield  {author} {\bibinfo {author} {\bibfnamefont {L.~D.}\ \bibnamefont
  {Contreras-Pulido}}, \bibinfo {author} {\bibfnamefont {J.}~\bibnamefont
  {Splettstoesser}}, \bibinfo {author} {\bibfnamefont {M.}~\bibnamefont
  {Governale}}, \bibinfo {author} {\bibfnamefont {J.}~\bibnamefont
  {K{\"o}nig}}, \ and\ \bibinfo {author} {\bibfnamefont {M.}~\bibnamefont
  {B{\"u}ttiker}},\ }\href {\doibase 10.1103/PhysRevB.85.075301} {\bibfield
  {journal} {\bibinfo  {journal} {Phys. Rev. B}\ }\textbf {\bibinfo {volume}
  {85}},\ \bibinfo {pages} {075301} (\bibinfo {year} {2012})}\BibitemShut
  {NoStop}%
\bibitem [{\citenamefont {Fulton}\ and\ \citenamefont
  {Dolan}(1987)}]{fulton_observation_1987}%
  \BibitemOpen
  \bibfield  {author} {\bibinfo {author} {\bibfnamefont {T.~A.}\ \bibnamefont
  {Fulton}}\ and\ \bibinfo {author} {\bibfnamefont {G.~J.}\ \bibnamefont
  {Dolan}},\ }\href {\doibase 10.1103/PhysRevLett.59.109} {\bibfield  {journal}
  {\bibinfo  {journal} {Phys. Rev. Lett.}\ }\textbf {\bibinfo {volume} {59}},\
  \bibinfo {pages} {109} (\bibinfo {year} {1987})}\BibitemShut {NoStop}%
\bibitem [{\citenamefont {Nitzan}\ and\ \citenamefont
  {Ratner}(2003)}]{nitzan_electron_2003}%
  \BibitemOpen
  \bibfield  {author} {\bibinfo {author} {\bibfnamefont {A.}~\bibnamefont
  {Nitzan}}\ and\ \bibinfo {author} {\bibfnamefont {M.~A.}\ \bibnamefont
  {Ratner}},\ }\href {\doibase 10.1126/science.1081572} {\bibfield  {journal}
  {\bibinfo  {journal} {Science}\ }\textbf {\bibinfo {volume} {300}},\ \bibinfo
  {pages} {1384} (\bibinfo {year} {2003})}\BibitemShut {NoStop}%
\bibitem [{\citenamefont {Segal}\ and\ \citenamefont
  {Nitzan}(2002)}]{segal_heating_2002}%
  \BibitemOpen
  \bibfield  {author} {\bibinfo {author} {\bibfnamefont {D.}~\bibnamefont
  {Segal}}\ and\ \bibinfo {author} {\bibfnamefont {A.}~\bibnamefont {Nitzan}},\
  }\href {\doibase 10.1063/1.1495845} {\bibfield  {journal} {\bibinfo
  {journal} {The Journal of Chemical Physics}\ }\textbf {\bibinfo {volume}
  {117}},\ \bibinfo {pages} {3915} (\bibinfo {year} {2002})}\BibitemShut
  {NoStop}%
\bibitem [{\citenamefont {Segal}\ \emph {et~al.}(2003)\citenamefont {Segal},
  \citenamefont {Nitzan},\ and\ \citenamefont
  {H{\"a}nggi}}]{segal_thermal_2003}%
  \BibitemOpen
  \bibfield  {author} {\bibinfo {author} {\bibfnamefont {D.}~\bibnamefont
  {Segal}}, \bibinfo {author} {\bibfnamefont {A.}~\bibnamefont {Nitzan}}, \
  and\ \bibinfo {author} {\bibfnamefont {P.}~\bibnamefont {H{\"a}nggi}},\
  }\href {\doibase 10.1063/1.1603211} {\bibfield  {journal} {\bibinfo
  {journal} {The Journal of Chemical Physics}\ }\textbf {\bibinfo {volume}
  {119}},\ \bibinfo {pages} {6840} (\bibinfo {year} {2003})}\BibitemShut
  {NoStop}%
\bibitem [{\citenamefont {Velizhanin}\ \emph {et~al.}(2008)\citenamefont
  {Velizhanin}, \citenamefont {Wang},\ and\ \citenamefont
  {Thoss}}]{velizhanin_heat_2008}%
  \BibitemOpen
  \bibfield  {author} {\bibinfo {author} {\bibfnamefont {K.~A.}\ \bibnamefont
  {Velizhanin}}, \bibinfo {author} {\bibfnamefont {H.}~\bibnamefont {Wang}}, \
  and\ \bibinfo {author} {\bibfnamefont {M.}~\bibnamefont {Thoss}},\ }\href
  {\doibase 10.1016/j.cplett.2008.05.065} {\bibfield  {journal} {\bibinfo
  {journal} {Chem. Phys. Lett.}\ }\textbf {\bibinfo {volume} {460}},\ \bibinfo
  {pages} {325} (\bibinfo {year} {2008})}\BibitemShut {NoStop}%
\bibitem [{\citenamefont {Chu}(2002)}]{chu_cold_2002}%
  \BibitemOpen
  \bibfield  {author} {\bibinfo {author} {\bibfnamefont {S.}~\bibnamefont
  {Chu}},\ }\href {\doibase 10.1038/416206a} {\bibfield  {journal} {\bibinfo
  {journal} {Nature}\ }\textbf {\bibinfo {volume} {416}},\ \bibinfo {pages}
  {206} (\bibinfo {year} {2002})}\BibitemShut {NoStop}%
\bibitem [{\citenamefont {K{\"o}hl}\ \emph {et~al.}(2005)\citenamefont
  {K{\"o}hl}, \citenamefont {Moritz}, \citenamefont {St{\"o}ferle},
  \citenamefont {G{\"u}nter},\ and\ \citenamefont
  {Esslinger}}]{kohl_fermionic_2005}%
  \BibitemOpen
  \bibfield  {author} {\bibinfo {author} {\bibfnamefont {M.}~\bibnamefont
  {K{\"o}hl}}, \bibinfo {author} {\bibfnamefont {H.}~\bibnamefont {Moritz}},
  \bibinfo {author} {\bibfnamefont {T.}~\bibnamefont {St{\"o}ferle}}, \bibinfo
  {author} {\bibfnamefont {K.}~\bibnamefont {G{\"u}nter}}, \ and\ \bibinfo
  {author} {\bibfnamefont {T.}~\bibnamefont {Esslinger}},\ }\href {\doibase
  10.1103/PhysRevLett.94.080403} {\bibfield  {journal} {\bibinfo  {journal}
  {Phys. Rev. Lett.}\ }\textbf {\bibinfo {volume} {94}},\ \bibinfo {pages}
  {080403} (\bibinfo {year} {2005})}\BibitemShut {NoStop}%
\bibitem [{\citenamefont {Ponomarev}\ \emph {et~al.}(2006)\citenamefont
  {Ponomarev}, \citenamefont {Madro{\~n}ero}, \citenamefont {Kolovsky},\ and\
  \citenamefont {Buchleitner}}]{ponomarev_atomic_2006}%
  \BibitemOpen
  \bibfield  {author} {\bibinfo {author} {\bibfnamefont {A.~V.}\ \bibnamefont
  {Ponomarev}}, \bibinfo {author} {\bibfnamefont {J.}~\bibnamefont
  {Madro{\~n}ero}}, \bibinfo {author} {\bibfnamefont {A.~R.}\ \bibnamefont
  {Kolovsky}}, \ and\ \bibinfo {author} {\bibfnamefont {A.}~\bibnamefont
  {Buchleitner}},\ }\href {\doibase 10.1103/PhysRevLett.96.050404} {\bibfield
  {journal} {\bibinfo  {journal} {Phys. Rev. Lett.}\ }\textbf {\bibinfo
  {volume} {96}},\ \bibinfo {pages} {050404} (\bibinfo {year}
  {2006})}\BibitemShut {NoStop}%
\bibitem [{\citenamefont {Bloch}\ \emph {et~al.}(2008)\citenamefont {Bloch},
  \citenamefont {Dalibard},\ and\ \citenamefont
  {Zwerger}}]{bloch_many-body_2008}%
  \BibitemOpen
  \bibfield  {author} {\bibinfo {author} {\bibfnamefont {I.}~\bibnamefont
  {Bloch}}, \bibinfo {author} {\bibfnamefont {J.}~\bibnamefont {Dalibard}}, \
  and\ \bibinfo {author} {\bibfnamefont {W.}~\bibnamefont {Zwerger}},\ }\href
  {http://journals.aps.org/rmp/abstract/10.1103/RevModPhys.80.885} {\bibfield
  {journal} {\bibinfo  {journal} {Rev. Mod. Phys.}\ }\textbf {\bibinfo {volume}
  {80}},\ \bibinfo {pages} {885} (\bibinfo {year} {2008})}\BibitemShut
  {NoStop}%
\bibitem [{\citenamefont {Schneider}\ \emph {et~al.}(2012)\citenamefont
  {Schneider}, \citenamefont {Porras},\ and\ \citenamefont
  {Schaetz}}]{schneider_experimental_2012}%
  \BibitemOpen
  \bibfield  {author} {\bibinfo {author} {\bibfnamefont {C.}~\bibnamefont
  {Schneider}}, \bibinfo {author} {\bibfnamefont {D.}~\bibnamefont {Porras}}, \
  and\ \bibinfo {author} {\bibfnamefont {T.}~\bibnamefont {Schaetz}},\ }\href
  {\doibase 10.1088/0034-4885/75/2/024401} {\bibfield  {journal} {\bibinfo
  {journal} {Rep. Prog. Phys.}\ }\textbf {\bibinfo {volume} {75}},\ \bibinfo
  {pages} {024401} (\bibinfo {year} {2012})}\BibitemShut {NoStop}%
\bibitem [{\citenamefont {Alicki}(1976)}]{alicki_detailed_1976}%
  \BibitemOpen
  \bibfield  {author} {\bibinfo {author} {\bibfnamefont {R.}~\bibnamefont
  {Alicki}},\ }\href {\doibase 10.1016/0034-4877(76)90046-X} {\bibfield
  {journal} {\bibinfo  {journal} {Rep. Math. Phys.}\ }\textbf {\bibinfo
  {volume} {10}},\ \bibinfo {pages} {249} (\bibinfo {year} {1976})}\BibitemShut
  {NoStop}%
\bibitem [{\citenamefont {Alicki}(1979)}]{alicki_quantum_1979}%
  \BibitemOpen
  \bibfield  {author} {\bibinfo {author} {\bibfnamefont {R.}~\bibnamefont
  {Alicki}},\ }\href {\doibase 10.1088/0305-4470/12/5/007} {\bibfield
  {journal} {\bibinfo  {journal} {J. Phys. A: Math. Gen.}\ }\textbf {\bibinfo
  {volume} {12}},\ \bibinfo {pages} {L103} (\bibinfo {year}
  {1979})}\BibitemShut {NoStop}%
\bibitem [{\citenamefont {Alicki}(2010)}]{alicki_field-theoretical_2010}%
  \BibitemOpen
  \bibfield  {author} {\bibinfo {author} {\bibfnamefont {R.}~\bibnamefont
  {Alicki}},\ }in\ \href
  {http://link.springer.com/chapter/10.1007/978-3-642-11914-9_5} {\emph
  {\bibinfo {booktitle} {Quantum {Information}, {Computation} and
  {Cryptography}}}},\ \bibinfo {series and number} {\bibinfo {series} {Lecture
  {Notes} in {Physics}}\ No.\ \bibinfo {number} {808}},\ \bibinfo {editor}
  {edited by\ \bibinfo {editor} {\bibfnamefont {F.}~\bibnamefont {Benatti}},
  \bibinfo {editor} {\bibfnamefont {M.}~\bibnamefont {Fannes}}, \bibinfo
  {editor} {\bibfnamefont {R.}~\bibnamefont {Floreanini}}, \ and\ \bibinfo
  {editor} {\bibfnamefont {D.}~\bibnamefont {Petritis}}}\ (\bibinfo
  {publisher} {Springer Berlin Heidelberg},\ \bibinfo {year} {2010})\ pp.\
  \bibinfo {pages} {151--174}\BibitemShut {NoStop}%
\bibitem [{\citenamefont {Alicki}\ and\ \citenamefont
  {Fannes}(2001)}]{alicki_quantum_2001}%
  \BibitemOpen
  \bibfield  {author} {\bibinfo {author} {\bibfnamefont {R.}~\bibnamefont
  {Alicki}}\ and\ \bibinfo {author} {\bibfnamefont {M.}~\bibnamefont
  {Fannes}},\ }\href@noop {} {{\selectlanguage {en}\emph {\bibinfo {title}
  {Quantum {Dynamical} {Systems}}}}}\ (\bibinfo  {publisher} {Oxford University
  Press},\ \bibinfo {year} {2001})\BibitemShut {NoStop}%
\bibitem [{\citenamefont {Benatti}\ \emph {et~al.}(2010)\citenamefont
  {Benatti}, \citenamefont {Fannes}, \citenamefont {Floreanini},\ and\
  \citenamefont {Petritis}}]{benatti_quantum_2010}%
  \BibitemOpen
  \bibfield  {author} {\bibinfo {author} {\bibfnamefont {F.}~\bibnamefont
  {Benatti}}, \bibinfo {author} {\bibfnamefont {M.}~\bibnamefont {Fannes}},
  \bibinfo {author} {\bibfnamefont {R.}~\bibnamefont {Floreanini}}, \ and\
  \bibinfo {author} {\bibfnamefont {D.}~\bibnamefont {Petritis}},\ }\href@noop
  {} {{\selectlanguage {en}\emph {\bibinfo {title} {Quantum {Information},
  {Computation} and {Cryptography}: {An} {Introductory} {Survey} of {Theory},
  {Technology} and {Experiments}}}}}\ (\bibinfo  {publisher} {Springer Science
  \& Business Media},\ \bibinfo {year} {2010})\BibitemShut {NoStop}%
\bibitem [{\citenamefont {Bratteli}\ and\ \citenamefont
  {Robinson}(1997)}]{bratteli_operator_1997}%
  \BibitemOpen
  \bibfield  {author} {\bibinfo {author} {\bibfnamefont {O.}~\bibnamefont
  {Bratteli}}\ and\ \bibinfo {author} {\bibfnamefont {D.~W.}\ \bibnamefont
  {Robinson}},\ }\href {http://dx.doi.org/10.1007/978-3-662-03444-6}
  {{\selectlanguage {English}\emph {\bibinfo {title} {Operator algebras and
  quantum statistical mechanics equilibrium states. {Models} in quantum
  statistical mechanics}}}}\ (\bibinfo  {publisher} {Springer},\ \bibinfo
  {address} {Berlin},\ \bibinfo {year} {1997})\BibitemShut {NoStop}%
\bibitem [{Note1()}]{Note1}%
  \BibitemOpen
  \bibinfo {note} {A $C^*$-algebra $\protect \mathcal {A}$ is by definition
  equipped with a norm which fulfils the properties $\delimiter 69645069 x^*
  \delimiter 86422285 = \delimiter 69645069 x \delimiter 86422285 $ and
  $\delimiter 69645069 x^* x\delimiter 86422285 = \delimiter 69645069
  x^*\delimiter 86422285 \delimiter 69645069 x\delimiter 86422285 $ for all $x
  \in \protect \mathcal {A}$. Here, this demand is strong enough to fix the
  norm in a unique way which is why it is referred to as the $C^*$-norm. For a
  much more complete and formal introduction to $C^*$-algebras, see for example
  \cite {bratteli_operator_1987}.}\BibitemShut {Stop}%
\bibitem [{\citenamefont {Davies}\ and\ \citenamefont
  {Lewis}(1970)}]{davies_operational_1970}%
  \BibitemOpen
  \bibfield  {author} {\bibinfo {author} {\bibfnamefont {E.~B.}\ \bibnamefont
  {Davies}}\ and\ \bibinfo {author} {\bibfnamefont {J.~T.}\ \bibnamefont
  {Lewis}},\ }\href {\doibase 10.1007/BF01647093} {\bibfield  {journal}
  {\bibinfo  {journal} {Commun. Math. Phys.}\ }\textbf {\bibinfo {volume}
  {17}},\ \bibinfo {pages} {239} (\bibinfo {year} {1970})}\BibitemShut
  {NoStop}%
\bibitem [{\citenamefont {Holevo}(2001)}]{holevo_statistical_2001}%
  \BibitemOpen
  \bibfield  {author} {\bibinfo {author} {\bibfnamefont {A.~S.}\ \bibnamefont
  {Holevo}},\ }\href@noop {} {{\selectlanguage {en}\emph {\bibinfo {title}
  {Statistical {Structure} of {Quantum} {Theory}}}}}\ (\bibinfo  {publisher}
  {Springer Science \& Business Media},\ \bibinfo {year} {2001})\BibitemShut
  {NoStop}%
\bibitem [{\citenamefont {Maassen}(2010)}]{maassen_quantum_2010}%
  \BibitemOpen
  \bibfield  {author} {\bibinfo {author} {\bibfnamefont {H.}~\bibnamefont
  {Maassen}},\ }in\ \href
  {http://link.springer.com/chapter/10.1007/978-3-642-11914-9_3}
  {{\selectlanguage {en}\emph {\bibinfo {booktitle} {Quantum {Information},
  {Computation} and {Cryptography}}}}},\ \bibinfo {series and number} {\bibinfo
  {series} {Lecture {Notes} in {Physics}}\ No.\ \bibinfo {number} {808}},\
  \bibinfo {editor} {edited by\ \bibinfo {editor} {\bibfnamefont
  {F.}~\bibnamefont {Benatti}}, \bibinfo {editor} {\bibfnamefont
  {M.}~\bibnamefont {Fannes}}, \bibinfo {editor} {\bibfnamefont
  {R.}~\bibnamefont {Floreanini}}, \ and\ \bibinfo {editor} {\bibfnamefont
  {D.}~\bibnamefont {Petritis}}}\ (\bibinfo  {publisher} {Springer Berlin
  Heidelberg},\ \bibinfo {year} {2010})\ pp.\ \bibinfo {pages} {65--108},\
  \bibinfo {note} {dOI: 10.1007/978-3-642-11914-9\_3}\BibitemShut {NoStop}%
\bibitem [{\citenamefont {Bratteli}\ and\ \citenamefont
  {Robinson}(1987)}]{bratteli_operator_1987}%
  \BibitemOpen
  \bibfield  {author} {\bibinfo {author} {\bibfnamefont {O.}~\bibnamefont
  {Bratteli}}\ and\ \bibinfo {author} {\bibfnamefont {D.~W.}\ \bibnamefont
  {Robinson}},\ }\href {http://link.springer.com/10.1007/978-3-662-02520-8}
  {\emph {\bibinfo {title} {Operator {Algebras} and {Quantum} {Statistical}
  {Mechanics} 1}}}\ (\bibinfo  {publisher} {Springer Berlin Heidelberg},\
  \bibinfo {address} {Berlin, Heidelberg},\ \bibinfo {year} {1987})\BibitemShut
  {NoStop}%
\bibitem [{\citenamefont {Verbeure}(2011)}]{verbeure_many-body_2011}%
  \BibitemOpen
  \bibfield  {author} {\bibinfo {author} {\bibfnamefont {A.}~\bibnamefont
  {Verbeure}},\ }\href@noop {} {\emph {\bibinfo {title} {Many-body boson
  systems: half a century later}}},\ Theoretical and mathematical physics\
  (\bibinfo  {publisher} {Springer},\ \bibinfo {address} {London ; New York},\
  \bibinfo {year} {2011})\BibitemShut {NoStop}%
\bibitem [{\citenamefont {Bardeen}\ \emph {et~al.}(1957)\citenamefont
  {Bardeen}, \citenamefont {Cooper},\ and\ \citenamefont
  {Schrieffer}}]{PhysRev.108.1175}%
  \BibitemOpen
  \bibfield  {author} {\bibinfo {author} {\bibfnamefont {J.}~\bibnamefont
  {Bardeen}}, \bibinfo {author} {\bibfnamefont {L.~N.}\ \bibnamefont {Cooper}},
  \ and\ \bibinfo {author} {\bibfnamefont {J.~R.}\ \bibnamefont {Schrieffer}},\
  }\href {\doibase 10.1103/PhysRev.108.1175} {\bibfield  {journal} {\bibinfo
  {journal} {Phys. Rev.}\ }\textbf {\bibinfo {volume} {108}},\ \bibinfo {pages}
  {1175} (\bibinfo {year} {1957})}\BibitemShut {NoStop}%
\bibitem [{\citenamefont {Bogolubov}(1960)}]{Bogolubov1960S1}%
  \BibitemOpen
  \bibfield  {author} {\bibinfo {author} {\bibfnamefont {N.}~\bibnamefont
  {Bogolubov}},\ }\href {\doibase
  http://dx.doi.org/10.1016/0031-8914(60)90196-8} {\bibfield  {journal}
  {\bibinfo  {journal} {Physica}\ }\textbf {\bibinfo {volume} {26, Supplement
  1}},\ \bibinfo {pages} {S1 } (\bibinfo {year} {1960})}\BibitemShut {NoStop}%
\bibitem [{\citenamefont {Haag}(1962)}]{Haag1962}%
  \BibitemOpen
  \bibfield  {author} {\bibinfo {author} {\bibfnamefont {R.}~\bibnamefont
  {Haag}},\ }\href {\doibase 10.1007/BF02731446} {\bibfield  {journal}
  {\bibinfo  {journal} {Il Nuovo Cimento (1955-1965)}\ }\textbf {\bibinfo
  {volume} {25}},\ \bibinfo {pages} {287} (\bibinfo {year} {1962})}\BibitemShut
  {NoStop}%
\bibitem [{\citenamefont {Emch}\ and\ \citenamefont
  {Guenin}(1966)}]{doi:10.1063/1.1931227}%
  \BibitemOpen
  \bibfield  {author} {\bibinfo {author} {\bibfnamefont {G.}~\bibnamefont
  {Emch}}\ and\ \bibinfo {author} {\bibfnamefont {M.}~\bibnamefont {Guenin}},\
  }\href {\doibase 10.1063/1.1931227} {\bibfield  {journal} {\bibinfo
  {journal} {Journal of Mathematical Physics}\ }\textbf {\bibinfo {volume}
  {7}},\ \bibinfo {pages} {915} (\bibinfo {year} {1966})},\ \Eprint
  {http://arxiv.org/abs/http://dx.doi.org/10.1063/1.1931227}
  {http://dx.doi.org/10.1063/1.1931227} \BibitemShut {NoStop}%
\bibitem [{\citenamefont {Thirring}\ and\ \citenamefont
  {Wehrl}(1967)}]{Thirring1967}%
  \BibitemOpen
  \bibfield  {author} {\bibinfo {author} {\bibfnamefont {W.}~\bibnamefont
  {Thirring}}\ and\ \bibinfo {author} {\bibfnamefont {A.}~\bibnamefont
  {Wehrl}},\ }\href {\doibase 10.1007/BF01653644} {\bibfield  {journal}
  {\bibinfo  {journal} {Communications in Mathematical Physics}\ }\textbf
  {\bibinfo {volume} {4}},\ \bibinfo {pages} {303} (\bibinfo {year}
  {1967})}\BibitemShut {NoStop}%
\bibitem [{\citenamefont {Thirring}(1968)}]{Thirring1968}%
  \BibitemOpen
  \bibfield  {author} {\bibinfo {author} {\bibfnamefont {W.}~\bibnamefont
  {Thirring}},\ }\href {\doibase 10.1007/BF01645661} {\bibfield  {journal}
  {\bibinfo  {journal} {Communications in Mathematical Physics}\ }\textbf
  {\bibinfo {volume} {7}},\ \bibinfo {pages} {181} (\bibinfo {year}
  {1968})}\BibitemShut {NoStop}%
\bibitem [{\citenamefont {Balslev}\ and\ \citenamefont
  {Verbeure}(1968)}]{Balslev1968}%
  \BibitemOpen
  \bibfield  {author} {\bibinfo {author} {\bibfnamefont {E.}~\bibnamefont
  {Balslev}}\ and\ \bibinfo {author} {\bibfnamefont {A.}~\bibnamefont
  {Verbeure}},\ }\href {\doibase 10.1007/BF01651218} {\bibfield  {journal}
  {\bibinfo  {journal} {Communications in Mathematical Physics}\ }\textbf
  {\bibinfo {volume} {7}},\ \bibinfo {pages} {55} (\bibinfo {year}
  {1968})}\BibitemShut {NoStop}%
\bibitem [{\citenamefont {Goderis}\ \emph {et~al.}(1991)\citenamefont
  {Goderis}, \citenamefont {Verbeure},\ and\ \citenamefont
  {Vets}}]{Goderis1991}%
  \BibitemOpen
  \bibfield  {author} {\bibinfo {author} {\bibfnamefont {D.}~\bibnamefont
  {Goderis}}, \bibinfo {author} {\bibfnamefont {A.}~\bibnamefont {Verbeure}}, \
  and\ \bibinfo {author} {\bibfnamefont {P.}~\bibnamefont {Vets}},\ }\href
  {\doibase 10.1007/BF02725688} {\bibfield  {journal} {\bibinfo  {journal} {Il
  Nuovo Cimento B (1971-1996)}\ }\textbf {\bibinfo {volume} {106}},\ \bibinfo
  {pages} {375} (\bibinfo {year} {1991})}\BibitemShut {NoStop}%
\bibitem [{Note2()}]{Note2}%
  \BibitemOpen
  \bibinfo {note} {An accessible introduction to fermionic quasi-free states
  can be found in~\cite {dierckx_fermionic_2008}.}\BibitemShut {Stop}%
\bibitem [{Note3()}]{Note3}%
  \BibitemOpen
  \bibinfo {note} {For an operator algebra of observables which acts on a
  Hilbert space, a state is said to be {\protect \em normal} if it can be
  represented by a density operator which is a trace-class operator on the same
  Hilbert space. However, because we consider abstract C*-algebras, it only
  makes sense to refer to {\protect \em normal} states in the context of a
  specific representation. Throughout our contribution we will always refer to
  {\protect \em normal} states as states which can be represented by a density
  matrix on the {\protect \em Fock space}, in other words, states which are
  normal with respect to the Fock representation. Note that in this
  representation the abstract operators $c$ are represented by the operators
  $a$ of (\ref {eq:CAR1}).}\BibitemShut {Stop}%
\bibitem [{\citenamefont {Pechukas}(1994)}]{pechukas_reduced_1994}%
  \BibitemOpen
  \bibfield  {author} {\bibinfo {author} {\bibfnamefont {P.}~\bibnamefont
  {Pechukas}},\ }\href {\doibase 10.1103/PhysRevLett.73.1060} {\bibfield
  {journal} {\bibinfo  {journal} {Phys. Rev. Lett.}\ }\textbf {\bibinfo
  {volume} {73}},\ \bibinfo {pages} {1060} (\bibinfo {year}
  {1994})}\BibitemShut {NoStop}%
\bibitem [{\citenamefont {Alicki}(1995)}]{alicki_comment_1995}%
  \BibitemOpen
  \bibfield  {author} {\bibinfo {author} {\bibfnamefont {R.}~\bibnamefont
  {Alicki}},\ }\href {\doibase 10.1103/PhysRevLett.75.3020} {\bibfield
  {journal} {\bibinfo  {journal} {Phys. Rev. Lett.}\ }\textbf {\bibinfo
  {volume} {75}},\ \bibinfo {pages} {3020} (\bibinfo {year}
  {1995})}\BibitemShut {NoStop}%
\bibitem [{\citenamefont {Pechukas}(1995)}]{pechukas_pechukas_1995}%
  \BibitemOpen
  \bibfield  {author} {\bibinfo {author} {\bibfnamefont {P.}~\bibnamefont
  {Pechukas}},\ }\href {\doibase 10.1103/PhysRevLett.75.3021} {\bibfield
  {journal} {\bibinfo  {journal} {Phys. Rev. Lett.}\ }\textbf {\bibinfo
  {volume} {75}},\ \bibinfo {pages} {3021} (\bibinfo {year}
  {1995})}\BibitemShut {NoStop}%
\bibitem [{\citenamefont {Stinespring}(1955)}]{stinespring_positive_1955}%
  \BibitemOpen
  \bibfield  {author} {\bibinfo {author} {\bibfnamefont {W.}~\bibnamefont
  {Stinespring}},\ }\href@noop {} {\bibfield  {journal} {\bibinfo  {journal}
  {Proc. Amer. Math. Soc.}\ }\bibinfo {series} {B},\ \textbf {\bibinfo {volume}
  {6}},\ \bibinfo {pages} {211} (\bibinfo {year} {1955})}\BibitemShut {NoStop}%
\bibitem [{\citenamefont {Kraus}(1971)}]{kraus_general_1971}%
  \BibitemOpen
  \bibfield  {author} {\bibinfo {author} {\bibfnamefont {K.}~\bibnamefont
  {Kraus}},\ }\href {\doibase 10.1016/0003-4916(71)90108-4} {\bibfield
  {journal} {\bibinfo  {journal} {Ann. Phys.}\ }\textbf {\bibinfo {volume}
  {64}},\ \bibinfo {pages} {311} (\bibinfo {year} {1971})}\BibitemShut
  {NoStop}%
\bibitem [{\citenamefont {Lindblad}(1976)}]{lindblad_generators_1976}%
  \BibitemOpen
  \bibfield  {author} {\bibinfo {author} {\bibfnamefont {G.}~\bibnamefont
  {Lindblad}},\ }\href {\doibase 10.1007/BF01608499} {\bibfield  {journal}
  {\bibinfo  {journal} {Commun. Math. Phys.}\ }\textbf {\bibinfo {volume}
  {48}},\ \bibinfo {pages} {119} (\bibinfo {year} {1976})}\BibitemShut
  {NoStop}%
\bibitem [{\citenamefont {Davies}(1976)}]{davies_quantum_1976}%
  \BibitemOpen
  \bibfield  {author} {\bibinfo {author} {\bibfnamefont {E.~B.}\ \bibnamefont
  {Davies}},\ }\href@noop {} {\emph {\bibinfo {title} {Quantum theory of open
  systems}}}\ (\bibinfo  {publisher} {Academic Press},\ \bibinfo {address}
  {London ; New York},\ \bibinfo {year} {1976})\BibitemShut {NoStop}%
\bibitem [{\citenamefont {Alicki}(1987)}]{alicki_quantum_1987}%
  \BibitemOpen
  \bibfield  {author} {\bibinfo {author} {\bibfnamefont {R.}~\bibnamefont
  {Alicki}},\ }\href@noop {} {{\selectlanguage {en}\emph {\bibinfo {title}
  {Quantum {Dynamical} {Semigroups} and {Applications}}}}},\ \bibinfo {edition}
  {2nd}\ ed.\ (\bibinfo  {publisher} {Springer Science \& Business Media},\
  \bibinfo {year} {1987})\BibitemShut {NoStop}%
\bibitem [{\citenamefont {Breuer}\ and\ \citenamefont
  {Petruccione}(2007)}]{breuer_theory_2007}%
  \BibitemOpen
  \bibfield  {author} {\bibinfo {author} {\bibfnamefont {H.-P.}\ \bibnamefont
  {Breuer}}\ and\ \bibinfo {author} {\bibfnamefont {F.}~\bibnamefont
  {Petruccione}},\ }\href@noop {} {{\selectlanguage {en}\emph {\bibinfo {title}
  {The {Theory} of {Open} {Quantum} {Systems}}}}}\ (\bibinfo  {publisher} {OUP
  Oxford},\ \bibinfo {year} {2007})\BibitemShut {NoStop}%
\bibitem [{\citenamefont {Gorini}\ \emph {et~al.}(1976)\citenamefont {Gorini},
  \citenamefont {Kossakowski},\ and\ \citenamefont
  {Sudarshan}}]{gorini_completely_1976}%
  \BibitemOpen
  \bibfield  {author} {\bibinfo {author} {\bibfnamefont {V.}~\bibnamefont
  {Gorini}}, \bibinfo {author} {\bibfnamefont {A.}~\bibnamefont {Kossakowski}},
  \ and\ \bibinfo {author} {\bibfnamefont {E.~C.~G.}\ \bibnamefont
  {Sudarshan}},\ }\href {\doibase 10.1063/1.522979} {\bibfield  {journal}
  {\bibinfo  {journal} {J. Math. Phys.}\ }\textbf {\bibinfo {volume} {17}},\
  \bibinfo {pages} {821} (\bibinfo {year} {1976})}\BibitemShut {NoStop}%
\bibitem [{\citenamefont {Davies}(1977)}]{davies_irreversible_1977}%
  \BibitemOpen
  \bibfield  {author} {\bibinfo {author} {\bibfnamefont {E.~B.}\ \bibnamefont
  {Davies}},\ }\href {\doibase 10.1007/BF01614549} {\bibfield  {journal}
  {\bibinfo  {journal} {Commun. Math. Phys.}\ }\textbf {\bibinfo {volume}
  {55}},\ \bibinfo {pages} {231} (\bibinfo {year} {1977})}\BibitemShut
  {NoStop}%
\bibitem [{\citenamefont {Alicki}(1978)}]{alicki_theory_1978}%
  \BibitemOpen
  \bibfield  {author} {\bibinfo {author} {\bibfnamefont {R.}~\bibnamefont
  {Alicki}},\ }\href {\doibase 10.1016/0034-4877(78)90030-7} {\bibfield
  {journal} {\bibinfo  {journal} {Rep. Math. Phys.}\ }\textbf {\bibinfo
  {volume} {14}},\ \bibinfo {pages} {27} (\bibinfo {year} {1978})}\BibitemShut
  {NoStop}%
\bibitem [{\citenamefont {Fannes}\ and\ \citenamefont
  {Verbeure}(1977{\natexlab{a}})}]{fannes_correlation_1977}%
  \BibitemOpen
  \bibfield  {author} {\bibinfo {author} {\bibfnamefont {M.}~\bibnamefont
  {Fannes}}\ and\ \bibinfo {author} {\bibfnamefont {A.}~\bibnamefont
  {Verbeure}},\ }\href {\doibase 10.1007/BF01626515} {\bibfield  {journal}
  {\bibinfo  {journal} {Commun. Math. Phys.}\ }\textbf {\bibinfo {volume}
  {55}},\ \bibinfo {pages} {125} (\bibinfo {year}
  {1977}{\natexlab{a}})}\BibitemShut {NoStop}%
\bibitem [{\citenamefont {Fannes}\ and\ \citenamefont
  {Verbeure}(1977{\natexlab{b}})}]{fannes_correlation_1977-1}%
  \BibitemOpen
  \bibfield  {author} {\bibinfo {author} {\bibfnamefont {M.}~\bibnamefont
  {Fannes}}\ and\ \bibinfo {author} {\bibfnamefont {A.}~\bibnamefont
  {Verbeure}},\ }\href {\doibase 10.1007/BF01625775} {\bibfield  {journal}
  {\bibinfo  {journal} {Commun. Math. Phys.}\ }\textbf {\bibinfo {volume}
  {57}},\ \bibinfo {pages} {165} (\bibinfo {year}
  {1977}{\natexlab{b}})}\BibitemShut {NoStop}%
\bibitem [{\citenamefont {Petz}\ and\ \citenamefont
  {Toth}(1993)}]{petz_bogoliubov_1993}%
  \BibitemOpen
  \bibfield  {author} {\bibinfo {author} {\bibfnamefont {D.}~\bibnamefont
  {Petz}}\ and\ \bibinfo {author} {\bibfnamefont {G.}~\bibnamefont {Toth}},\
  }\href {\doibase 10.1007/BF00739578} {\bibfield  {journal} {\bibinfo
  {journal} {Lett Math Phys}\ }\textbf {\bibinfo {volume} {27}},\ \bibinfo
  {pages} {205} (\bibinfo {year} {1993})}\BibitemShut {NoStop}%
\bibitem [{\citenamefont {Mehta}(2004)}]{mehta_random_2004}%
  \BibitemOpen
  \bibfield  {author} {\bibinfo {author} {\bibfnamefont {M.~L.}\ \bibnamefont
  {Mehta}},\ }\href@noop {} {{\selectlanguage {en}\emph {\bibinfo {title}
  {Random {Matrices}}}}}\ (\bibinfo  {publisher} {Elsevier/Academic press},\
  \bibinfo {address} {Amsterdam},\ \bibinfo {year} {2004})\BibitemShut
  {NoStop}%
\bibitem [{\citenamefont {Wishart}(1928)}]{wishart_generalised_1928}%
  \BibitemOpen
  \bibfield  {author} {\bibinfo {author} {\bibfnamefont {J.}~\bibnamefont
  {Wishart}},\ }\href {\doibase 10.1093/biomet/20A.1-2.32} {\bibfield
  {journal} {\bibinfo  {journal} {Biometrika}\ }\textbf {\bibinfo {volume}
  {20A}},\ \bibinfo {pages} {32} (\bibinfo {year} {1928})}\BibitemShut
  {NoStop}%
\bibitem [{Note4()}]{Note4}%
  \BibitemOpen
  \bibinfo {note} {From~(\ref {eq:FormalJ2}) it directly follows that attempts
  to saturate the bound by optimising $H$ are futile in the limit where
  $\lambda = 0.$}\BibitemShut {NoStop}%
\bibitem [{\citenamefont {Pedersen}(1989)}]{pedersen_analysis_1989}%
  \BibitemOpen
  \bibfield  {author} {\bibinfo {author} {\bibfnamefont {G.~K.}\ \bibnamefont
  {Pedersen}},\ }\href {http://dx.doi.org/10.1007/978-1-4612-1007-8}
  {{\selectlanguage {English}\emph {\bibinfo {title} {Analysis {Now}}}}}\
  (\bibinfo  {publisher} {Springer New York},\ \bibinfo {address} {New York,
  NY},\ \bibinfo {year} {1989})\BibitemShut {NoStop}%
\bibitem [{Note5()}]{Note5}%
  \BibitemOpen
  \bibinfo {note} {The Haar measure may be interpreted as the uniform
  distribution over the set of unitary matrices.}\BibitemShut {Stop}%
\bibitem [{\citenamefont {Mosonyi}\ \emph {et~al.}(2008)\citenamefont
  {Mosonyi}, \citenamefont {Hiai}, \citenamefont {Ogawa},\ and\ \citenamefont
  {Fannes}}]{mosonyi_asymptotic_2008}%
  \BibitemOpen
  \bibfield  {author} {\bibinfo {author} {\bibfnamefont {M.}~\bibnamefont
  {Mosonyi}}, \bibinfo {author} {\bibfnamefont {F.}~\bibnamefont {Hiai}},
  \bibinfo {author} {\bibfnamefont {T.}~\bibnamefont {Ogawa}}, \ and\ \bibinfo
  {author} {\bibfnamefont {M.}~\bibnamefont {Fannes}},\ }\href {\doibase
  10.1063/1.2953473} {\bibfield  {journal} {\bibinfo  {journal} {Journal of
  Mathematical Physics}\ }\textbf {\bibinfo {volume} {49}},\ \bibinfo {pages}
  {072104} (\bibinfo {year} {2008})}\BibitemShut {NoStop}%
\bibitem [{Note6()}]{Note6}%
  \BibitemOpen
  \bibinfo {note} {The integrability of a matrix-valued function can be
  understood in a component-wise way.}\BibitemShut {Stop}%
\bibitem [{\citenamefont {Petz}(1990)}]{petz_invitation_1990}%
  \BibitemOpen
  \bibfield  {author} {\bibinfo {author} {\bibfnamefont {D.}~\bibnamefont
  {Petz}},\ }\href@noop {} {{\selectlanguage {eng}\emph {\bibinfo {title} {An
  invitation to the algebra of canonical comutation relations}}}},\ \bibinfo
  {series} {Leuven notes in mathematical and theoretical physics {Series} {A}}\
  No.~\bibinfo {number} {2}\ (\bibinfo  {publisher} {Leuven Univ. Press},\
  \bibinfo {address} {Leuven},\ \bibinfo {year} {1990})\BibitemShut {NoStop}%
\bibitem [{Note7()}]{Note7}%
  \BibitemOpen
  \bibinfo {note} {We define this equation for all linear operators on Fock
  space, a set which we denote ${\protect \rm Lin}{\setbox \z@ \hbox
  {\frozen@everymath \@emptytoks \mathsurround \z@ $\nulldelimiterspace \z@
  \mathopen {}\mathclose \bgroup \originalleft (\vcenter to\@ne \big@size
  {}\aftergroup \egroup \originalright .$}\box \z@ }\Gamma (\protect \mathcal
  {H}){\setbox \z@ \hbox {\frozen@everymath \@emptytoks \mathsurround \z@
  $\nulldelimiterspace \z@ \mathopen {}\mathclose \bgroup \originalleft
  )\vcenter to\@ne \big@size {}\aftergroup \egroup \originalright .$}\box \z@
  }$.}\BibitemShut {Stop}%
\bibitem [{Note8()}]{Note8}%
  \BibitemOpen
  \bibinfo {note} {What is quite obvious from a physical point of view, think,
  \protect \emph {e.g.,\ }of a micro-maser scenario \cite
  {lugiato_connection_1987} where a single quantised mode of the radiation
  field is pumped by a sequence of two-level atoms which enter the resonator in
  the excited state and interact resonantly with the resonator mode. The steady
  state is there defined by an equilibrium of gain and loss (damping) of the
  resonator mode. Note, however, that in this paradigmatic realisation the
  effective pump rate depends nonlinearly on the occupation number of the
  resonator mode.}\BibitemShut {Stop}%
\bibitem [{\citenamefont {Christandl}\ \emph {et~al.}(2005)\citenamefont
  {Christandl}, \citenamefont {Datta}, \citenamefont {Dorlas}, \citenamefont
  {Ekert}, \citenamefont {Kay},\ and\ \citenamefont
  {Landahl}}]{christandl_perfect_2005}%
  \BibitemOpen
  \bibfield  {author} {\bibinfo {author} {\bibfnamefont {M.}~\bibnamefont
  {Christandl}}, \bibinfo {author} {\bibfnamefont {N.}~\bibnamefont {Datta}},
  \bibinfo {author} {\bibfnamefont {T.~C.}\ \bibnamefont {Dorlas}}, \bibinfo
  {author} {\bibfnamefont {A.}~\bibnamefont {Ekert}}, \bibinfo {author}
  {\bibfnamefont {A.}~\bibnamefont {Kay}}, \ and\ \bibinfo {author}
  {\bibfnamefont {A.~J.}\ \bibnamefont {Landahl}},\ }\href {\doibase
  10.1103/PhysRevA.71.032312} {\bibfield  {journal} {\bibinfo  {journal} {Phys.
  Rev. A}\ }\textbf {\bibinfo {volume} {71}},\ \bibinfo {pages} {032312}
  (\bibinfo {year} {2005})}\BibitemShut {NoStop}%
\bibitem [{\citenamefont {Zech}\ \emph {et~al.}(2014)\citenamefont {Zech},
  \citenamefont {Mulet}, \citenamefont {Wellens},\ and\ \citenamefont
  {Buchleitner}}]{zech_centrosymmetry_2014}%
  \BibitemOpen
  \bibfield  {author} {\bibinfo {author} {\bibfnamefont {T.}~\bibnamefont
  {Zech}}, \bibinfo {author} {\bibfnamefont {R.}~\bibnamefont {Mulet}},
  \bibinfo {author} {\bibfnamefont {T.}~\bibnamefont {Wellens}}, \ and\
  \bibinfo {author} {\bibfnamefont {A.}~\bibnamefont {Buchleitner}},\ }\href
  {\doibase 10.1088/1367-2630/16/5/055002} {\bibfield  {journal} {\bibinfo
  {journal} {New J. Phys.}\ }\textbf {\bibinfo {volume} {16}},\ \bibinfo
  {pages} {055002} (\bibinfo {year} {2014})}\BibitemShut {NoStop}%
\bibitem [{\citenamefont {Walschaers}\ \emph {et~al.}(2015)\citenamefont
  {Walschaers}, \citenamefont {Mulet}, \citenamefont {Wellens},\ and\
  \citenamefont {Buchleitner}}]{walschaers_statistical_2015}%
  \BibitemOpen
  \bibfield  {author} {\bibinfo {author} {\bibfnamefont {M.}~\bibnamefont
  {Walschaers}}, \bibinfo {author} {\bibfnamefont {R.}~\bibnamefont {Mulet}},
  \bibinfo {author} {\bibfnamefont {T.}~\bibnamefont {Wellens}}, \ and\
  \bibinfo {author} {\bibfnamefont {A.}~\bibnamefont {Buchleitner}},\ }\href
  {\doibase 10.1103/PhysRevE.91.042137} {\bibfield  {journal} {\bibinfo
  {journal} {Phys. Rev. E}\ }\textbf {\bibinfo {volume} {91}},\ \bibinfo
  {pages} {042137} (\bibinfo {year} {2015})}\BibitemShut {NoStop}%
\bibitem [{\citenamefont {Ortega}\ \emph {et~al.}(2016)\citenamefont {Ortega},
  \citenamefont {Stegmann},\ and\ \citenamefont
  {Benet}}]{ortega_efficient_2016}%
  \BibitemOpen
  \bibfield  {author} {\bibinfo {author} {\bibfnamefont {A.}~\bibnamefont
  {Ortega}}, \bibinfo {author} {\bibfnamefont {T.}~\bibnamefont {Stegmann}}, \
  and\ \bibinfo {author} {\bibfnamefont {L.}~\bibnamefont {Benet}},\ }\href
  {http://arxiv.org/abs/1605.01445} {\bibfield  {journal} {\bibinfo  {journal}
  {arXiv:1605.01445 [cond-mat, physics:quant-ph]}\ } (\bibinfo {year}
  {2016})},\ \bibinfo {note} {arXiv: 1605.01445}\BibitemShut {NoStop}%
\bibitem [{\citenamefont {Landauer}(1957)}]{landauer_spatial_1957}%
  \BibitemOpen
  \bibfield  {author} {\bibinfo {author} {\bibfnamefont {R.}~\bibnamefont
  {Landauer}},\ }\href {\doibase 10.1147/rd.13.0223} {\bibfield  {journal}
  {\bibinfo  {journal} {IBM Journal of Research and Development}\ }\textbf
  {\bibinfo {volume} {1}},\ \bibinfo {pages} {223} (\bibinfo {year}
  {1957})}\BibitemShut {NoStop}%
\bibitem [{\citenamefont {Beenakker}(1997)}]{beenakker_random-matrix_1997}%
  \BibitemOpen
  \bibfield  {author} {\bibinfo {author} {\bibfnamefont {C.~W.~J.}\
  \bibnamefont {Beenakker}},\ }\href {\doibase 10.1103/RevModPhys.69.731}
  {\bibfield  {journal} {\bibinfo  {journal} {Rev. Mod. Phys.}\ }\textbf
  {\bibinfo {volume} {69}},\ \bibinfo {pages} {731} (\bibinfo {year}
  {1997})}\BibitemShut {NoStop}%
\bibitem [{\citenamefont {Schlawin}\ \emph {et~al.}(2012)\citenamefont
  {Schlawin}, \citenamefont {Cherroret},\ and\ \citenamefont
  {Buchleitner}}]{schlawin_bunching_2012}%
  \BibitemOpen
  \bibfield  {author} {\bibinfo {author} {\bibfnamefont {F.}~\bibnamefont
  {Schlawin}}, \bibinfo {author} {\bibfnamefont {N.}~\bibnamefont {Cherroret}},
  \ and\ \bibinfo {author} {\bibfnamefont {A.}~\bibnamefont {Buchleitner}},\
  }\href {\doibase 10.1209/0295-5075/99/14001} {\bibfield  {journal} {\bibinfo
  {journal} {EPL}\ }\textbf {\bibinfo {volume} {99}},\ \bibinfo {pages} {14001}
  (\bibinfo {year} {2012})}\BibitemShut {NoStop}%
\bibitem [{\citenamefont {Kropf}\ \emph {et~al.}(2016)\citenamefont {Kropf},
  \citenamefont {Gneiting},\ and\ \citenamefont
  {Buchleitner}}]{kropf_effective_2016}%
  \BibitemOpen
  \bibfield  {author} {\bibinfo {author} {\bibfnamefont {C.~M.}\ \bibnamefont
  {Kropf}}, \bibinfo {author} {\bibfnamefont {C.}~\bibnamefont {Gneiting}}, \
  and\ \bibinfo {author} {\bibfnamefont {A.}~\bibnamefont {Buchleitner}},\
  }\href {\doibase 10.1103/PhysRevX.6.031023} {\bibfield  {journal} {\bibinfo
  {journal} {Phys. Rev. X}\ }\textbf {\bibinfo {volume} {6}},\ \bibinfo {pages}
  {031023} (\bibinfo {year} {2016})}\BibitemShut {NoStop}%
\bibitem [{\citenamefont {Uzdin}\ \emph {et~al.}(2015)\citenamefont {Uzdin},
  \citenamefont {Levy},\ and\ \citenamefont
  {Kosloff}}]{uzdin_equivalence_2015}%
  \BibitemOpen
  \bibfield  {author} {\bibinfo {author} {\bibfnamefont {R.}~\bibnamefont
  {Uzdin}}, \bibinfo {author} {\bibfnamefont {A.}~\bibnamefont {Levy}}, \ and\
  \bibinfo {author} {\bibfnamefont {R.}~\bibnamefont {Kosloff}},\ }\href
  {\doibase 10.1103/PhysRevX.5.031044} {\bibfield  {journal} {\bibinfo
  {journal} {Phys. Rev. X}\ }\textbf {\bibinfo {volume} {5}},\ \bibinfo {pages}
  {031044} (\bibinfo {year} {2015})}\BibitemShut {NoStop}%
\bibitem [{\citenamefont {Uzdin}\ and\ \citenamefont
  {Kosloff}(2014)}]{uzdin_universal_2014}%
  \BibitemOpen
  \bibfield  {author} {\bibinfo {author} {\bibfnamefont {R.}~\bibnamefont
  {Uzdin}}\ and\ \bibinfo {author} {\bibfnamefont {R.}~\bibnamefont
  {Kosloff}},\ }\href {\doibase 10.1209/0295-5075/108/40001} {\bibfield
  {journal} {\bibinfo  {journal} {EPL}\ }\textbf {\bibinfo {volume} {108}},\
  \bibinfo {pages} {40001} (\bibinfo {year} {2014})}\BibitemShut {NoStop}%
\bibitem [{\citenamefont {Walschaers}(2016)}]{walschaers_efficient_2016}%
  \BibitemOpen
  \bibfield  {author} {\bibinfo {author} {\bibfnamefont {M.}~\bibnamefont
  {Walschaers}},\ }\emph {\bibinfo {title} {Efficient {Quantum} {Transport}}},\
  \href {https://www.freidok.uni-freiburg.de/data/11065} {\bibinfo {type}
  {{PhD} {Thesis}}},\ \bibinfo  {school} {Albert-Ludwigs Universit{\"a}t
  Freiburg \& KU Leuven}, \bibinfo {address} {Freiburg} (\bibinfo {year}
  {2016})\BibitemShut {NoStop}%
\bibitem [{\citenamefont {Wang}\ \emph {et~al.}(2015)\citenamefont {Wang},
  \citenamefont {Ren},\ and\ \citenamefont {Cao}}]{wang_nonequilibrium_2015}%
  \BibitemOpen
  \bibfield  {author} {\bibinfo {author} {\bibfnamefont {C.}~\bibnamefont
  {Wang}}, \bibinfo {author} {\bibfnamefont {J.}~\bibnamefont {Ren}}, \ and\
  \bibinfo {author} {\bibfnamefont {J.}~\bibnamefont {Cao}},\ }\href {\doibase
  10.1038/srep11787} {\bibfield  {journal} {\bibinfo  {journal} {Scientific
  Reports}\ }\textbf {\bibinfo {volume} {5}},\ \bibinfo {pages} {11787}
  (\bibinfo {year} {2015})}\BibitemShut {NoStop}%
\bibitem [{\citenamefont {Dierckx}\ \emph {et~al.}(2008)\citenamefont
  {Dierckx}, \citenamefont {Fannes},\ and\ \citenamefont
  {Pogorzelska}}]{dierckx_fermionic_2008}%
  \BibitemOpen
  \bibfield  {author} {\bibinfo {author} {\bibfnamefont {B.}~\bibnamefont
  {Dierckx}}, \bibinfo {author} {\bibfnamefont {M.}~\bibnamefont {Fannes}}, \
  and\ \bibinfo {author} {\bibfnamefont {M.}~\bibnamefont {Pogorzelska}},\
  }\href {\doibase 10.1063/1.2841326} {\bibfield  {journal} {\bibinfo
  {journal} {J. Math. Phys.}\ }\textbf {\bibinfo {volume} {49}},\ \bibinfo
  {pages} {032109} (\bibinfo {year} {2008})}\BibitemShut {NoStop}%
\bibitem [{\citenamefont {Lugiato}\ \emph {et~al.}(1987)\citenamefont
  {Lugiato}, \citenamefont {Scully},\ and\ \citenamefont
  {Walther}}]{lugiato_connection_1987}%
  \BibitemOpen
  \bibfield  {author} {\bibinfo {author} {\bibfnamefont {L.~A.}\ \bibnamefont
  {Lugiato}}, \bibinfo {author} {\bibfnamefont {M.~O.}\ \bibnamefont {Scully}},
  \ and\ \bibinfo {author} {\bibfnamefont {H.}~\bibnamefont {Walther}},\ }\href
  {\doibase 10.1103/PhysRevA.36.740} {\bibfield  {journal} {\bibinfo  {journal}
  {Phys. Rev. A}\ }\textbf {\bibinfo {volume} {36}},\ \bibinfo {pages} {740}
  (\bibinfo {year} {1987})}\BibitemShut {NoStop}%
\end{thebibliography}%

\end{document}